\documentclass[floatfix,showpacs]{revtex4}
\usepackage{epsfig}
\usepackage{graphicx}
\newcommand{\vphi}{\varphi}

\begin{document}

\title{ Gravitating Monopole--Antimonopole Chains \\
and Vortex Rings}

\author{Burkhard Kleihaus, Jutta Kunz and Yasha Shnir} 
\affiliation{Institut f\"ur Physik, Universit\"at Oldenburg,
D-26111, Oldenburg, Germany}


\begin{abstract}

We construct monopole-antimonopole chain
and vortex solutions in Yang-Mills-Higgs theory
coupled to Einstein gravity.
The solutions are static, axially symmetric and asymptotically flat.
They are characterized by two integers $(m,n)$
where $m$ is related to the polar angle and $n$ to the
azimuthal angle.
Solutions with $n=1$ and $n=2$ correspond to chains
of $m$ monopoles and antimonopoles. Here
the Higgs field vanishes at $m$ isolated points along the symmetry axis.
Larger values of $n$ give rise to vortex solutions,
where the Higgs field vanishes on one or more rings,
centered around the symmetry axis.
When gravity is coupled to the flat space solutions,
a branch of gravitating monopole-antimonopole chain
or vortex solutions arises, and merges 
at a maximal value of the coupling constant
with a second branch of solutions.
This upper branch has no flat space limit.
Instead in the limit of vanishing coupling constant
it either connects to a Bartnik-McKinnon or generalized
Bartnik-McKinnon solution,
or, for $m>4$, $n>4$, it connects to a new Einstein-Yang-Mills solution.
In this latter case further branches of solutions appear.
For small values of the coupling constant on the upper branches, 
the solutions correspond to composite systems,
consisting of a scaled inner Einstein-Yang-Mills solution
and an outer Yang-Mills-Higgs solution.

\end{abstract}
\maketitle

\section{Introduction}

The non-trivial vacuum structure of SU(2) Yang-Mills-Higgs (YMH) theory
allows for the existence of regular non-perturbative finite energy 
solutions, such as monopoles, multimonopoles and
monopole-antimonopole systems.
While spherically symmetric monopoles carry unit topological
charge \cite{mono,PraSom}, monopoles with charge $n>1$
are axially symmetric \cite{WeinbergGuth,RebbiRossi,mmono,KKT} 
or possess no rotational symmetry at all \cite{CorGod,monoDS}.

In the Bogomol'nyi-Prasad-Sommerfield (BPS) limit of vanishing Higgs potential,
the monopole and multimonopole solutions satisfy
a set of first order equations, the Bogomol'nyi equations \cite{Bogo}.
The spherically symmetric and axially symmetric BPS (multi)monopole solutions
are known analytically \cite{PraSom,mmono}.
In these solutions all nodes of the Higgs field are
superimposed at a single point.
In BPS multimonopole solutions with only discrete symmetries,
recently constructed numerically,
the nodes of the Higgs field can be located at several isolated points
\cite{monoDS}.
The energy of BPS solutions satisfies exactly the lower energy
bound given by the topological charge.
In particular, in the BPS limit
the repulsive and attractive forces between monopoles exactly compensate, 
thus BPS monopoles experience no net interaction \cite{Manton77}.

The configuration space of YMH theory consists of topological sectors
characterized by the topological charge of the Higgs field.
As shown by Taubes \cite{Taubes}, each topological sector contains
besides the (multi)monopole solutions further regular, finite energy solutions,
which do not satisfy the first order Bogomol'nyi equations, 
but only the set of second order field equations.
These solutions form saddlepoints of the energy functional
\cite{Taubes}.
The simplest such solution,
the mono\-pole--antimonopole pair solution \cite{Rueber,mapKK},
is topologically trivial. 
It possesses axial symmetry, and the two nodes of its
Higgs field are located symmetrically on the positive and negative
$z$-axis. 

Recently we have constructed axially symmetric saddlepoint solutions,
where the Higgs field vanishes at $m>2$ isolated points on the
symmetry axis.
These represent chains of single monopoles and antimonopoles
located on the symmetry axis in alternating order \cite{KKS}.
For an equal number of monopoles and antimonopoles, 
i.e.~for even $m$,
the chains reside in the topologically trivial sector.
When the number of monopoles exceeds the number of antimonopoles by one,
i.e.~for odd $m$,
the chains reside in the sector with topological charge one.
While such chains can also be formed from doubly charged
monopoles and antimonopoles ($n=2$), 
chains with higher charged monopoles ($n>2$) cannot be formed.
Instead completely different solutions appear, where the Higgs field
vanishes on one or more rings centered around the symmetry axis 
\cite{KKS}. We therefore refer to these solutions as vortex rings.

When gravity is coupled to YMH theory,
this has significant effect on the
monopole, multimonopole and monopole-antimonopole pair solutions.
When the coupling constant is increased from zero \cite{gmono,HKK},
a branch of gravitating monopole solutions emerges smoothly
from the flat space 't Hooft-Polyakov monopole solution.
The coupling constant $\alpha$,
entering the Einstein-Yang-Mills-Higgs (EYMH) equations,
is proportional to the Higgs vacuum expectation value $\eta$ 
and the square root of the gravitational constant $G$
and to the square of the Higgs vacuum expectation value $\eta$.
With increasing coupling constant $\alpha$
the mass of the gravitating monopole solutions decreases.
The branch of gravitating monopole solutions
extends up to a maximal value $\alpha_{\rm max}$,
beyond which gravity becomes too strong
for regular monopole solutions to persist \cite{gmono,HKK}.
For vanishing Higgs self-coupling constant, 
this first gravitating monopole branch merges with a second 
branch at $\alpha_{\rm max}$,
which extends slightly backwards,
until at a critical value $\alpha_{\rm cr}$ of the coupling constant
a degenerate horizon develops \cite{foot}.
The exterior space time of the solution then
corresponds to the one of an extremal Reissner-Nordstr\"om (RN)
black hole with unit magnetic charge \cite{gmono}.
At $\alpha_{\rm cr}$ this second branch of 
gravitating monopole solutions thus bifurcates with the branch of
extremal Reissner-Nordstr\"om (RN) solutions.
Remarkably,
the additional attraction in the YMH system due to the coupling to gravity 
also allows for bound monopoles, not present in flat space \cite{HKK}.

For the monopole-antimonopole pair solution, on the other hand,
we observe a different coupling constant dependence
for the gravitating solutions \cite{MAP}.
Again a branch of gravitating monopole-antimonopole pair solutions
emerges smoothly from the flat space solution,
and merges at a maximal value of the coupling constant $\alpha_{\rm max}$
with a second branch of gravitating monopole-antimonopole pair solutions.
This second branch, 
however, extends all the way back to vanishing coupling constant.
Along the first branch the mass of the solutions decreases with increasing 
$\alpha$, since with increasing gravitational strength
the attraction in the system increases.
Along the second branch the mass increases strongly
with decreasing coupling constant and diverges in the limit
of vanishing coupling constant.
This upper branch may be considered as being obtained
by decreasing the Higgs expectation value.
Along the upper branch the solutions shrink to zero size
in the limit $\alpha \rightarrow 0$.
Scaling the coordinates, the mass and the Higgs field by $\alpha$
shows, that the scaled solutions
approach the lowest mass Bartnik-McKinnon (BM) solution \cite{BM}
of Einstein-Yang-Mills (EYM) theory
in the limit $\alpha \rightarrow 0$.

Here we consider the effect of gravity on the
monopole-antimonopole chains and vortex rings \cite{KKS}.
Characterizing these solutions by 
the integers $m$ and $n$, related to the polar angle $\theta$ 
and the azimuthal angle $\varphi$, respectively,
we first consider chains, where $n \le 2$,
and then vortex rings, where $n \ge 3$.
When $m$ is even, these solutions reside in the
topologically trivial sector, thus we expect
a similar dependence on the coupling constant
as for the monopole-antimonopole pair.
When $m$ is odd, the solutions reside in the sector
with charge $n$, where the corresponding multimonopoles reside.
Therefore at first sight a connection with RN solutions appears 
to be possible, similar as observed for gravitating (multi)monopoles.
We find, however, that all these unstable gravitating solutions, 
chains and vortex rings alike,
show the same general coupling constant dependence
as observed for the monopole-antimonopole pair.
Two branches of solutions arise, a lower branch connected
to the flat space solution and an upper branch connected to 
an EYM solution. 
For $m \le 3$ these EYM solutions 
correspond to the spherically symmetric BM solutions ($n=1$) 
or their axially symmetric generalizations ($n>1$) \cite{BM,KK}.
For $m \ge 4$ new EYM solutions arise in the limit \cite{IKKS}, 
implying further branches of gravitating EYMH solutions.

In this paper we present
gravitating monopole-antimonopole chains and vortex rings
for vanishing Higgs self-coupling.
In section II we present the action, the axially
symmetric Ansatz and the boundary conditions.
In section III we discuss
the properties of these solutions
with particular emphasis on the limit $\alpha \rightarrow 0$,
and we present our conclusions in section IV.

\section{\bf Einstein-Yang-Mills-Higgs Solutions}

\subsection{\bf Einstein-Yang-Mills-Higgs action}

We consider static axially symmetric 
monopole-antimonopole chains and vortex rings
in SU(2) Einstein-Yang-Mills-Higgs 
theory with action 
\begin{equation} \label{action}
S =  \int \left\{ \frac{R}{16\pi G} 
-\frac{1}{2} {\rm Tr} 
\,\left( F_{\mu\nu}F^{\mu\nu} \right)
-\frac{1}{4} {\rm Tr}
\left(  D_\mu \Phi\, D^\mu \Phi  \right)
-\frac{\lambda}{8} 
 {\rm Tr} \left(\Phi^2 - \eta^2 \right)^2 
\right\}
\sqrt{- g} d^4 x
\end{equation}
with curvature scalar $R$,
$su(2)$ field strength tensor
\begin{equation}
F_{\mu\nu} = \partial_\mu A_\nu - \partial_\nu A_\mu + i e [A_\mu, A_\nu] \ ,
\end{equation}
gauge potential $A_\mu = A_\mu^a \tau^a/2$,
and covariant derivative of the Higgs field $\Phi = \Phi^a \tau^a$
in the adjoint representation
\begin{equation}
D_\mu \Phi = \partial_\mu \Phi +i e [A_\mu, \Phi] \ .
\end{equation}
Here $G$ and $e$ denote the gravitational and gauge coupling constants,
respectively,
$\eta$ denotes the vacuum expectation value of the Higgs field,
and $\lambda$ the strength of the Higgs self-coupling.

Under SU(2) gauge transformations $U$,
the gauge potentials transform as
\begin{equation}
A_{\mu}' = U A_{\mu} U^\dagger + \frac{i}{e} (\partial_\mu U) U^\dagger
\ , \label{gtgen} \end{equation}
and the Higgs field transforms as
\begin{equation}
\Phi' = U \Phi U^\dagger
\ . \label{gtgen2} \end{equation}

The nonzero vacuum expectation value of the Higgs field
breaks the non-Abelian SU(2) gauge symmetry to the Abelian U(1) symmetry.
The particle spectrum of the theory then consists of a massless photon,
two massive vector bosons of mass $M_v = e\eta$,
and a massive scalar field $M_s = {\sqrt {2 \lambda}}\, \eta$.
In the BPS limit the scalar field also becomes massless,
since $\lambda = 0$, i.e.~the Higgs potential vanishes.

Variation of the action (\ref{action}) with respect to the metric
$g^{\mu\nu}$ leads to the Einstein equations
\begin{equation}
G_{\mu\nu}= R_{\mu\nu}-\frac{1}{2}g_{\mu\nu}R = 8\pi G T_{\mu\nu}
\  \label{ee} \end{equation}
with stress-energy tensor
\begin{eqnarray}
T_{\mu\nu} &=& g_{\mu\nu}L_M -2 \frac{\partial L_M}{\partial g^{\mu\nu}}
 \nonumber \\
  &=&
      2\, {\rm Tr}\,
    ( F_{\mu\alpha} F_{\nu\beta} g^{\alpha\beta}
   -\frac{1}{4} g_{\mu\nu} F_{\alpha\beta} F^{\alpha\beta}) \nonumber \\
  &+&
      {\rm Tr}\, (\frac{1}{2}D_\mu \Phi D_\nu \Phi
    -\frac{1}{4} g_{\mu\nu} D_\alpha \Phi D^\alpha \Phi)
   -\frac{\lambda}{8}g_{\mu\nu} {\rm Tr}(\Phi^2 - \eta^2)^2
\ .
\end{eqnarray}

Variation with respect to the gauge field $A_\mu$
and the Higgs field $\Phi$
leads to the matter field equations,
\begin{eqnarray}
& &\frac{1}{\sqrt{-g}} D_\mu(\sqrt{-g} F^{\mu\nu})
   -\frac{1}{4} i e [\Phi, D^\nu \Phi ] = 0 \ ,
\label{feqA} \\
& & \frac{1}{\sqrt{-g}} D_\mu(\sqrt{-g} D^\mu \Phi)
+\lambda (\Phi^2 -\eta^2) \Phi  = 0 \ ,
\label{feqPhi}
\end{eqnarray}
respectively.

\subsection{\bf Static axially symmetric Ansatz}

To obtain gravitating static axially symmetric solutions,
we employ isotropic coordinates \cite{KK,HKK,IKKS}.
In terms of the spherical coordinates $r$, $\theta$ and $\vphi$
the isotropic metric reads
\begin{equation}
ds^2=
  - f dt^2 +  \frac{m}{f} d r^2 + \frac{m r^2}{f} d \theta^2
           +  \frac{l r^2 \sin^2 \theta}{f} d\vphi^2
\ , \label{metric2} \end{equation}
where the metric functions
$f$, $m$ and $l$ are functions of
the coordinates $r$ and $\theta$, only.
The $z$-axis ($\theta=0, \pi$) represents the symmetry axis.
Regularity on this axis requires \cite{book}
\begin{equation}
m|_{\theta=0, \pi}=l|_{\theta=0, \pi}
\ . \label{lm} \end{equation}

We parametrize the gauge potential and the Higgs field by the Ansatz
\cite{KKS}
\begin{eqnarray}
A_\mu dx^\mu
& = &
\left( \frac{K_1}{r} dr + (1-K_2)d\theta\right)\frac{\tau_\vphi^{(n)}}{2e}
\nonumber \\
&-& n \sin\theta \left( K_3\frac{\tau_r^{(n,m)}}{2e}
                     +(1-K_4)\frac{\tau_\theta^{(n,m)}}{2e}\right) d\vphi
\ , \label{ansatzA} \\
\Phi
& = &
\eta \left( \Phi_1\tau_r^{(n,m)}+ \Phi_2\tau_\theta^{(n,m)} \right) \  .
\label{ansatzPhi}
\end{eqnarray}
where the $su(2)$ matrices
$\tau_r^{(n,m)}$, $\tau_\theta^{(n,m)}$, and $\tau_\vphi^{(n)}$
are defined as products of the spatial unit vectors
\begin{eqnarray}
{\hat e}_r^{(n,m)} & = & \left(
\sin(m\theta) \cos(n\vphi), \sin(m\theta)\sin(n\vphi), \cos(m\theta)
\right)\ , \nonumber \\
{\hat e}_\theta^{(n,m)} & = & \left(
\cos(m\theta) \cos(n\vphi), \cos(m\theta)\sin(n\vphi), -\sin(m\theta)
\right)\ , \nonumber \\
{\hat e}_\vphi^{(n)} & = & \left( -\sin(n\vphi), \cos(n\vphi), 0 \right)\ ,
\label{unit_e}
\end{eqnarray}
with the Pauli matrices $\tau^a = (\tau_x, \tau_y, \tau_z)$, i.e.
\begin{eqnarray}
\tau_r^{(n,m)}  & = &
\sin(m\theta) \tau_\rho^{(n)} + \cos(m\theta) \tau_z \ ,
\nonumber\\
\tau_\theta^{(n,m)} & = &
\cos(m\theta) \tau_\rho^{(n)} - \sin(m\theta) \tau_z \ ,
\nonumber\\
\tau_\vphi^{(n)} & = &
 -\sin(n\vphi) \tau_x + \cos(n\vphi)\tau_y \ ,
\nonumber
\end{eqnarray}
with $\tau_\rho^{(n)} =\cos(n\vphi) \tau_x + \sin(n\vphi)\tau_y $.
For $m=2$, $n=1$ the Ansatz corresponds to the one for
the monopole-antimonopole pair solutions \cite{Rueber,mapKK,MAP},
while for $m=1$, $n>1$ it corresponds to
the Ansatz for axially symmetric multimonopoles \cite{RebbiRossi,KKT,HKK}.
The four gauge field functions $K_i$ and two Higgs field functions 
$\Phi_i$ depend on the coordinates $r$ and $\theta$, only.
All functions are even or odd w.r.t.~reflection symmetry, $z \rightarrow -z$.

The gauge transformation
\begin{equation}
U = \exp \{i \Gamma (r,\theta) \tau_\vphi^{(n)}/2\}
\  \end{equation}
leaves the Ansatz form-invariant \cite{BriKu}.
To construct regular solutions we have to fix the gauge \cite{KKT}.
Here we impose the gauge condition \cite{KKS}
\begin{equation}
 r \partial_r K_1 - \partial_\theta K_2 = 0
\ . \label{gauge} \end{equation}

With this Ansatz the equations of motion reduce to a set of 
9 coupled partial differential equations,
to be solved numerically subject to the set of boundary conditions,
discussed below.

\subsection{\bf Boundary conditions}

To obtain globally regular asymptotically flat solutions
with the proper symmetries,
we must impose appropriate boundary conditions \cite{HKK,KKS}.

{\sl Boundary conditions at the origin}

Regularity of the solutions at the origin ($r=0$) 
requires for the metric functions the boundary conditions 
\begin{equation}
\partial_r f(r,\theta)|_{r=0}= 
\partial_r m(r,\theta)|_{r=0}= 
\partial_r l(r,\theta)|_{r=0}= 0
\ , \label{bc2a} \end{equation}
whereas the gauge field functions $K_i$ satisfy
\begin{equation}
K_1(0,\theta)= K_3(0,\theta)= 0\ , \ \ \ \
K_2(0,\theta)= K_4(0,\theta)= 1 \ ,
\end{equation}
and the Higgs field functions $\Phi_i$ satisfy
\begin{equation}
\sin(m\theta) \Phi_1(0,\theta) + \cos(m\theta) \Phi_2(0,\theta) = 0 \ ,
\end{equation}
\begin{equation}
\left.\partial_r\left[\cos(m\theta) \Phi_1(r,\theta)
              - \sin(m\theta) \Phi_2(r,\theta)\right] \right|_{r=0} = 0 \ ,
\end{equation}
i.e.~$\Phi_\rho(0,\theta) =0$, $\partial_r \Phi_z(0,\theta) =0$.

{\sl Boundary conditions at infinity}

Asymptotic flatness imposes on the metric functions of the solutions
at infinity ($r=\infty$) the boundary conditions
\begin{equation}
f \longrightarrow 1 \ , \ \ \ 
m \longrightarrow 1 \ , \ \ \ 
l \longrightarrow 1 \ 
\ . \label{bc1a} \end{equation}
Considering the gauge field at infinity, we require that
solutions in the vacuum sector $Q=0$, where $m=2k$, tend to
a gauge transformed trivial solution,
$$
\Phi \ \longrightarrow \eta U \tau_z U^\dagger \   , \ \ \
A_\mu \ \longrightarrow  \ \frac{i}{e} (\partial_\mu U) U^\dagger \ ,
$$
and that solutions in the sector with topological charge $n$, where $m=2k+1$,
tend to
$$
\Phi  \longrightarrow  U \Phi_\infty^{(1,n)} U^\dagger \   , \ \ \
A_\mu \ \longrightarrow \ U A_{\mu \infty}^{(1,n)} U^\dagger
+\frac{i}{e} (\partial_\mu U) U^\dagger \  ,
$$
where
$$ \Phi_\infty^{(1,n)} =\eta \tau_r^{(1,n)}\ , \ \ \
A_{\mu \infty}^{(1,n)}dx^\mu =
\frac{\tau_\vphi^{(n)}}{2e} d\theta
- n\sin\theta \frac{\tau_\theta^{(1,n)}}{2e} d\vphi
$$
is the asymptotic solution of a charge $n$ multimonopole,
and $U = \exp\{-i k \theta\tau_\vphi^{(n)}\}$, both
for even and odd $m$.

In terms of the functions $K_1 - K_4$, $\Phi_1$, $\Phi_2$ these boundary
conditions read
\begin{equation}
K_1 \longrightarrow 0 \ , \ \ \ \
K_2 \longrightarrow 1 - m \ , \ \ \ \
\label{K12infty}
\end{equation}
\begin{equation}
K_3 \longrightarrow \frac{\cos\theta - \cos(m\theta)}{\sin\theta}
\ \ \ m \ {\rm odd} \ , \ \ \
K_3 \longrightarrow \frac{1 - \cos(m\theta)}{\sin\theta}
\ \ \ m \ {\rm even} \ , \ \ \
\label{K3infty}
\end{equation}
\begin{equation}
K_4 \longrightarrow 1- \frac{\sin(m\theta)}{\sin\theta} \ ,
\label{K4infty}
\end{equation}
\begin{equation}        \label{Phiinfty}
\Phi_1\longrightarrow  1 \ , \ \ \ \ \Phi_2 \longrightarrow 0 \ .
\end{equation}

{\sl Boundary conditions along the symmetry axis}

The boundary conditions along the $z$-axis
($\theta=0$ and $\theta=\pi $) are determined by the
symmetries.
The metric functions satisfy along the axis
\begin{eqnarray}
& &\partial_\theta f = \partial_\theta m =
   \partial_\theta l =0 \ ,
\label{bc4a}
\end{eqnarray}
whereas the matter field functions satisfy
\begin{equation}
K_1 = K_3 = \Phi_2 =0 \ , \ \ \  \
\partial_\theta K_2 = \partial_\theta K_4 = \partial_\theta \Phi_1 =0 \ .
\end{equation}

\section{Results and discussion}

We have constructed numerically gravitating chains and vortex rings,
subject to the above boundary conditions.
We first briefly address the numerical procedure,
and then present our results for the chains ($n=1$, 2)
and vortex rings ($n \ge 3$).
In particular, we consider the dependence of the solutions
on the value of the coupling constant $\alpha$,
and restrict to vanishing Higgs self-coupling $\lambda=0$.

\subsection{Numerical procedure}

To construct solutions subject to the above boundary conditions,
we map the infinite interval of the variable $r$
onto the unit interval of the compactified radial variable
$\bar x \in [0:1]$,
$$
\bar x = \frac{r}{1+r}
\ , $$
i.e., the partial derivative with respect to the radial coordinate
changes according to
$$
\partial_r \to (1- \bar x)^2\partial_{\bar x}
\ . $$

The numerical calculations are performed with the help of the FIDISOL package
based on the Newton-Raphson iterative procedure \cite{FIDI}.
It is therefore essential for the numerical procedure to have a reasonably good
initial configuration.
(For details see description and related documentation \cite{FIDI}.)
The equations are discretized on a non-equidistant grid in $x$ and $\theta$
with typical grids sizes of $70 \times 60$.
The estimates of the relative error for the functions
are of the order of $10^{-4}$, $10^{-3}$
and $10^{-2}$ for solutions with $m=2$, $m = 3,4$ and $m=5,6$, respectively.

\subsection{Dimensionless quantities}

Let us introduce the dimensionless coordinate $x$,
\begin{equation}
x= \eta e r
\ . \label{dimx} \end{equation}
The equations then
depend only on the dimensionless coupling constant $\alpha$,
\begin{equation}
\alpha^2 = 4\pi G\eta^2 \ 
\ . \end{equation}
The mass $M$ of the solutions
is related to the dimensionless mass $\mu$, via
\begin{equation}
\mu = \frac{e}{4\pi\eta} M
\ , \label{Mass2} \end{equation}
where $\mu$ is determined by the derivative of the metric function $f$
at infinity \cite{KK,HKK}
\begin{equation}
\mu = \frac{1}{2\alpha^2} \lim_{x \rightarrow \infty} x^2 \partial_x  f
\ . \label{mass} \end{equation}

\boldmath
\subsection{Gravitating chains: $n=1$}
\unboldmath

Let us consider first monopole-antimonopole chains
composed of poles of charge $n=1$.
The topological charge of these chains
is either zero (for even $m$) or unity (for odd $m$).
The monopole-antimonopole chains possess $m$ nodes 
of the Higgs field on the $z$-axis.
Because of reflection symmetry,
to each node on the positive $z$-axis
there corresponds a node on the negative $z$-axis.
For even $m$ ($m=2k$) the Higgs field does not have a node at the origin,
while for odd $m$ ($m=2k+1$) it does.
Associating with each pole the location of a single magnetic charge,
these chains then possess a total of $m$ magnetic monopoles
and antimonopoles, located in alternating order on the symmetry axis
\cite{KKS}.

The dependence of the $m=1$ solution, i.e.~of the 't Hooft-Polyakov monopole,
on gravity has been studied before.
When gravity is coupled, a branch of gravitating monopoles
emerges from the flat space solution and
extends up to a maximal value $\alpha_{\rm max}$,
where it merges with a second branch.
This second branch extends slightly backwards,
and bifurcates at a critical value $\alpha_{\rm cr}$ with the branch of
extremal RN solutions of unit charge.

The $m=2$ chain, i.e.~the monopole-antimonopole pair,
shows a different dependence on the coupling constant \cite{mapKK}.
Again, when gravity is coupled, a branch of gravitating 
monopole-antimonopole pair solutions emerges from the flat space solution,
and merges with a second branch of monopole-antimonopole pair solutions
at  a maximal value of the coupling constant $\alpha_{\rm max}$.
But since the monopole-antimonopole pair solutions reside in the
topologically trivial sector and thus carry no magnetic charge,
this upper branch of solutions
cannot bifurcate with a branch of extremal RN solutions
at a critical value $\alpha_{\rm cr}$ of the coupling constant.
Instead the upper branch extends all the way back to $\alpha = 0$.
Along the first branch the mass of the solutions decreases with increasing
$\alpha$, since with increasing gravitational strength
the attraction in the system increases.
Along the second branch, in contrast, the mass increases strongly
with decreasing $\alpha$, and the solutions shrink correspondingly.
In the limit of vanishing coupling constant the mass then diverges
and the solutions shrink to zero size.

Scaling the coordinates and the Higgs field, however, via
\begin{equation}
\hat{x} = \frac{x}{\alpha} \ , \ \ \
\hat{\Phi} = \alpha \Phi
\ , \end{equation}
leads to a limiting solution with finite size and 
finite scaled mass $\hat{\mu}$
\cite{mapKK},
\begin{equation}
{\hat{\mu}}= \alpha {\mu}
\ . \end{equation}
Indeed, after the scaling the field equations do not depend on $\alpha$.
Instead $\alpha$ appears in the 
asymptotic boundary conditions of the Higgs field,
$|\hat \Phi | \rightarrow \alpha$.
The Higgs field then becomes trivial
on the upper branch as $\alpha \rightarrow 0$.
Consequently, the scaled monopole-antimonopole solutions
approach an EYM solution as $\alpha \rightarrow 0$.
This limiting solution is the lowest mass BM solution \cite{BM}.
The mass and the scaled mass of the monopole-antimonopole pair solutions
are exhibited in Fig.~1.

\noindent\parbox{\textwidth}{
\centerline{
(a)\mbox{\epsfysize=6.0cm \epsffile{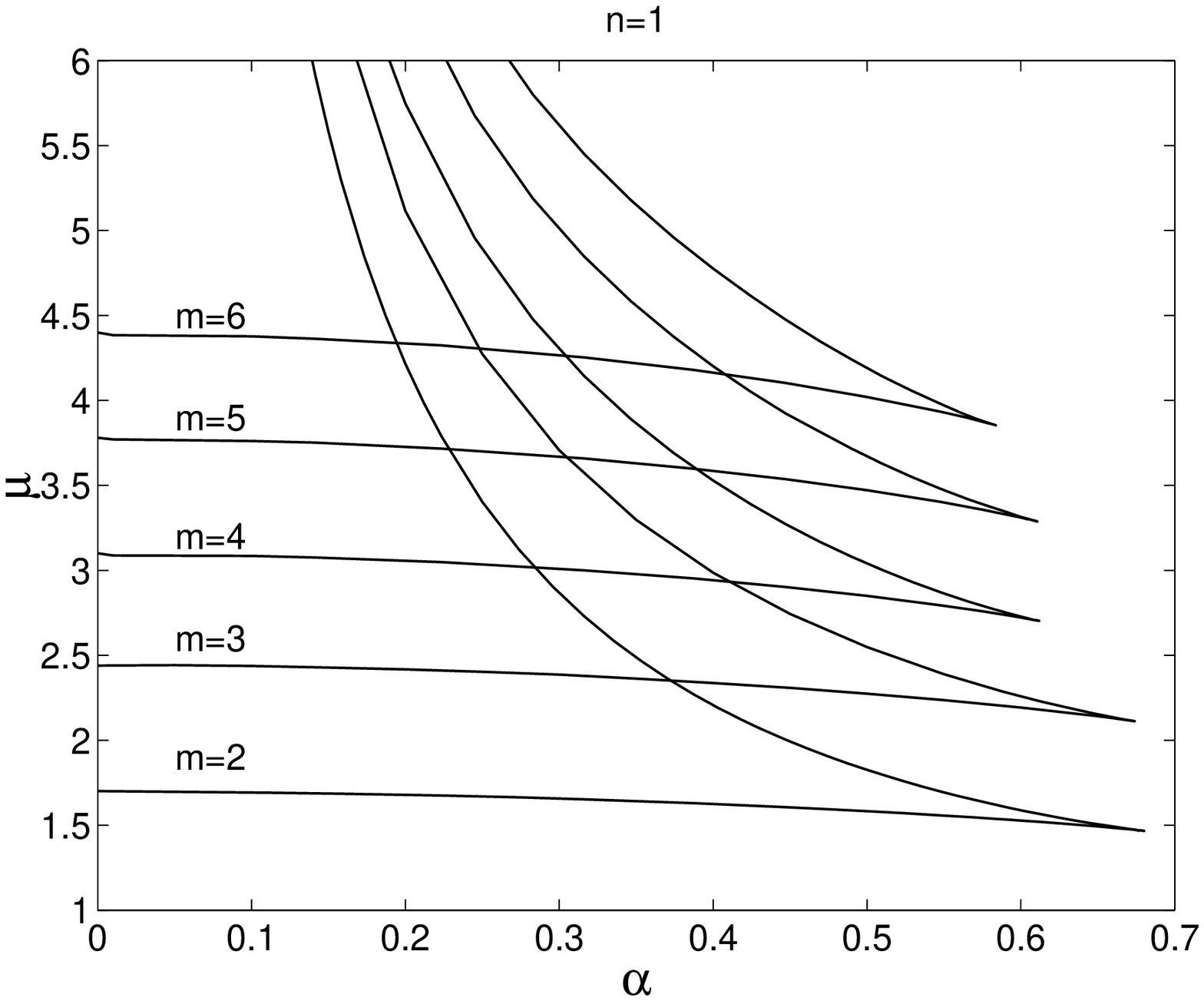} }
(b)\mbox{\epsfysize=6.0cm \epsffile{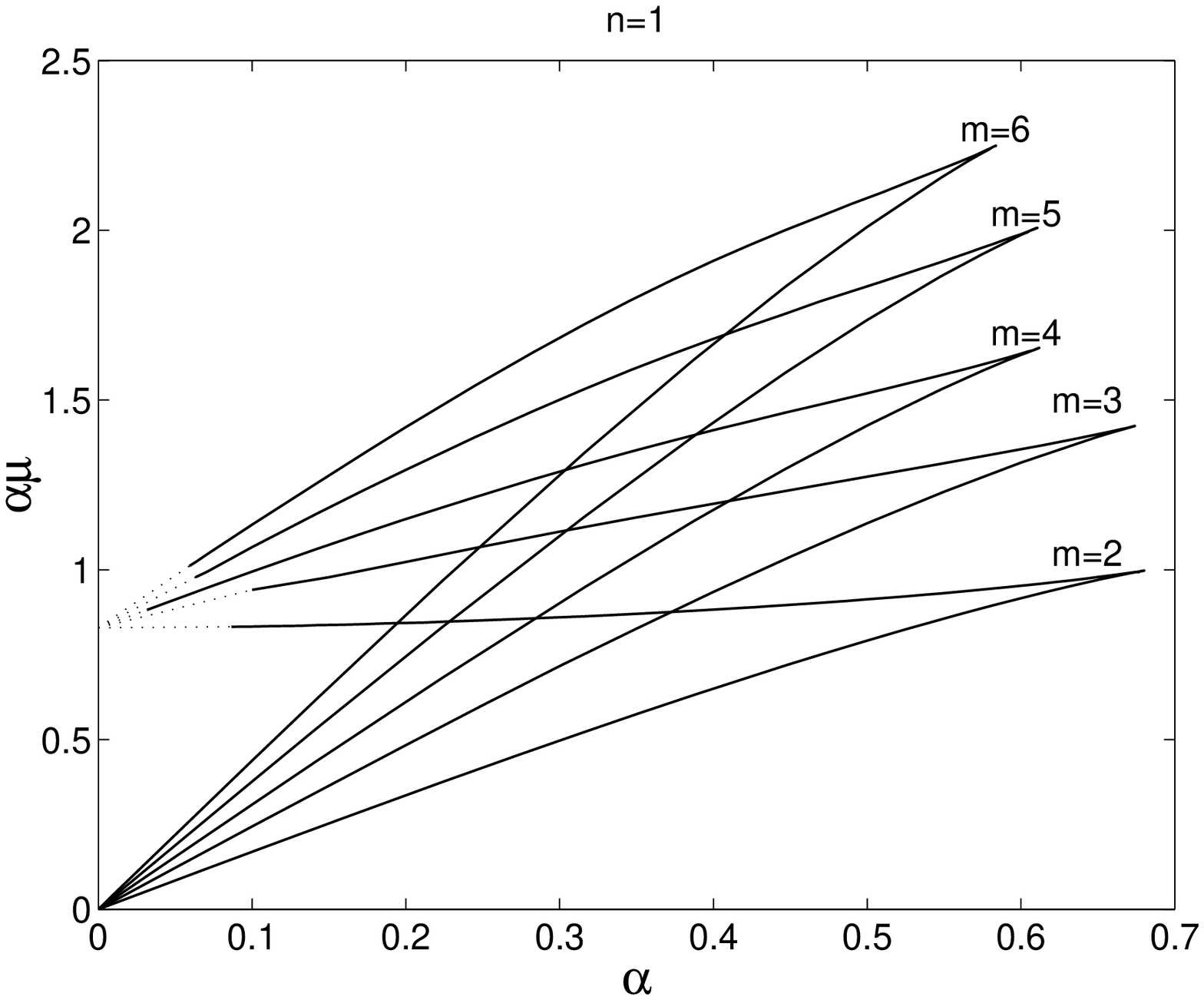} }
}\vspace{0.5cm}
{\bf Fig.~1} \small
The mass $\mu$ (a) and the scaled mass $\hat \mu$ (b)
are shown as functions of the coupling constant $\alpha$
for the chain solutions with $n=1$ and $m=2-6$.
The dotted lines extend the EYMH curves of the scaled mass to 
the mass of the lowest Bartnik-McKinnon solution.
\vspace{0.5cm}
}

Let us now consider chains with even $m$, $m>2$, 
which, like the $m=2$ monopole-antimonopole pair,
also reside in the vacuum sector.
As one may expect,
these chains exhibit an analogous dependence on the coupling constant $\alpha$
as the monopole-antimonopole pair. 
In the limit $\alpha \rightarrow 0$,
a lower branch of gravitating solutions emerges from the flat space solution, 
and merges at a maximal value $\alpha_{\rm max}$
with an upper branch of solutions, which extends all the way back to
$\alpha \rightarrow 0$. 
The mass of the chains with $m=4$ and $m=6$ is exhibited in Fig.~1a.
Here, for $\alpha \rightarrow 0$,
the emergence of the lower branches from the YMH limit is seen
as well as the divergence along the upper branches.
In Fig.~1b the scaled mass is exhibited. Clearly, the scaled mass
of these solutions approaches the mass of the lowest BM solution
in the limit $\alpha \rightarrow 0$ along the upper branch.
The values of the metric functions $f$ and $l$ at the origin
are shown in Fig.~2 for both branches of the chains with $m=4$ and $m=6$.
They evolve from unity in the flat space limit and reach
the corresponding values of the lowest mass BM solution, 
when $\alpha \rightarrow 0$ on the upper branch.

\noindent\parbox{\textwidth}{
\centerline{
(a)\mbox{\epsfysize=6.0cm \epsffile{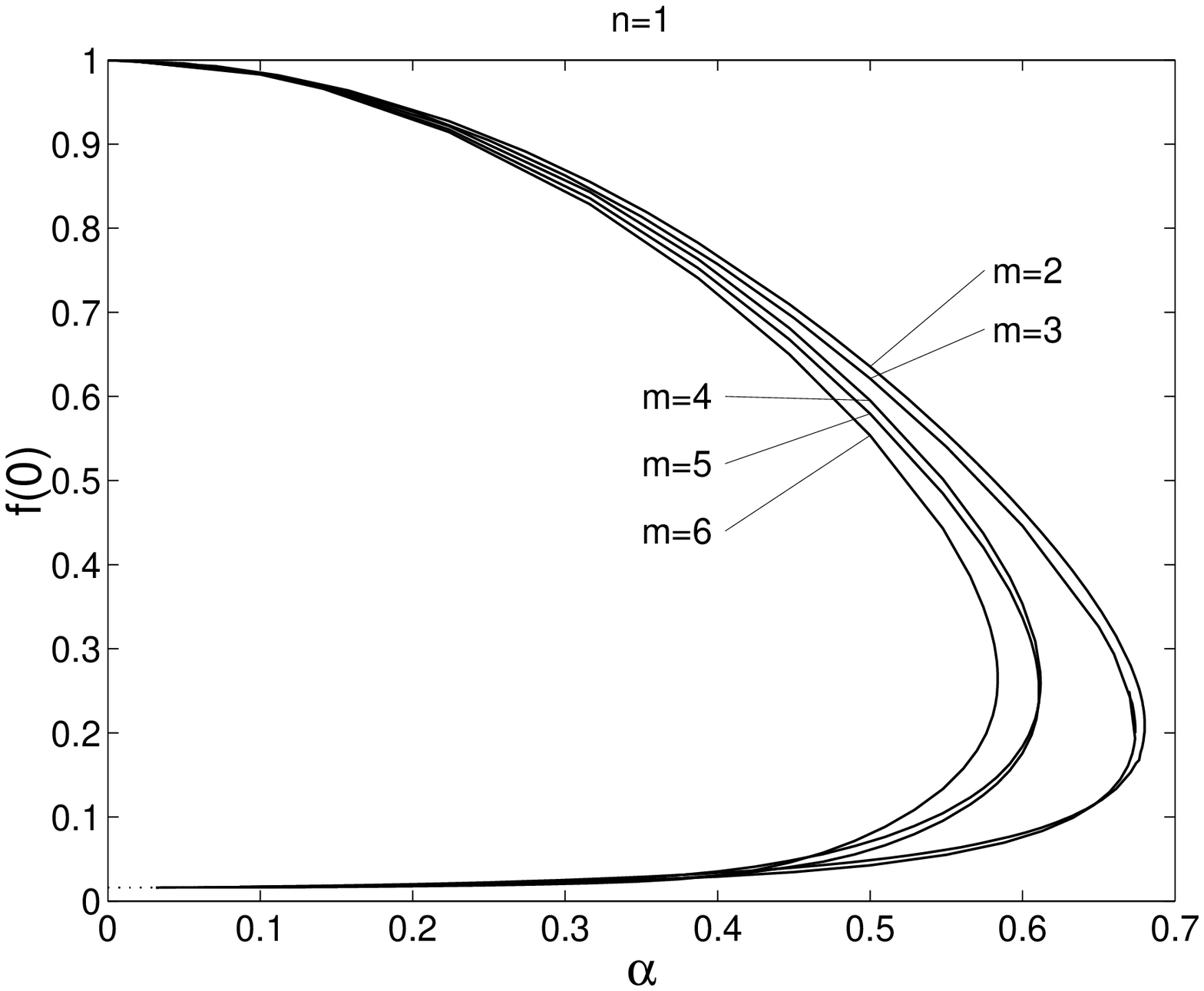} }
(b)\mbox{\epsfysize=6.0cm \epsffile{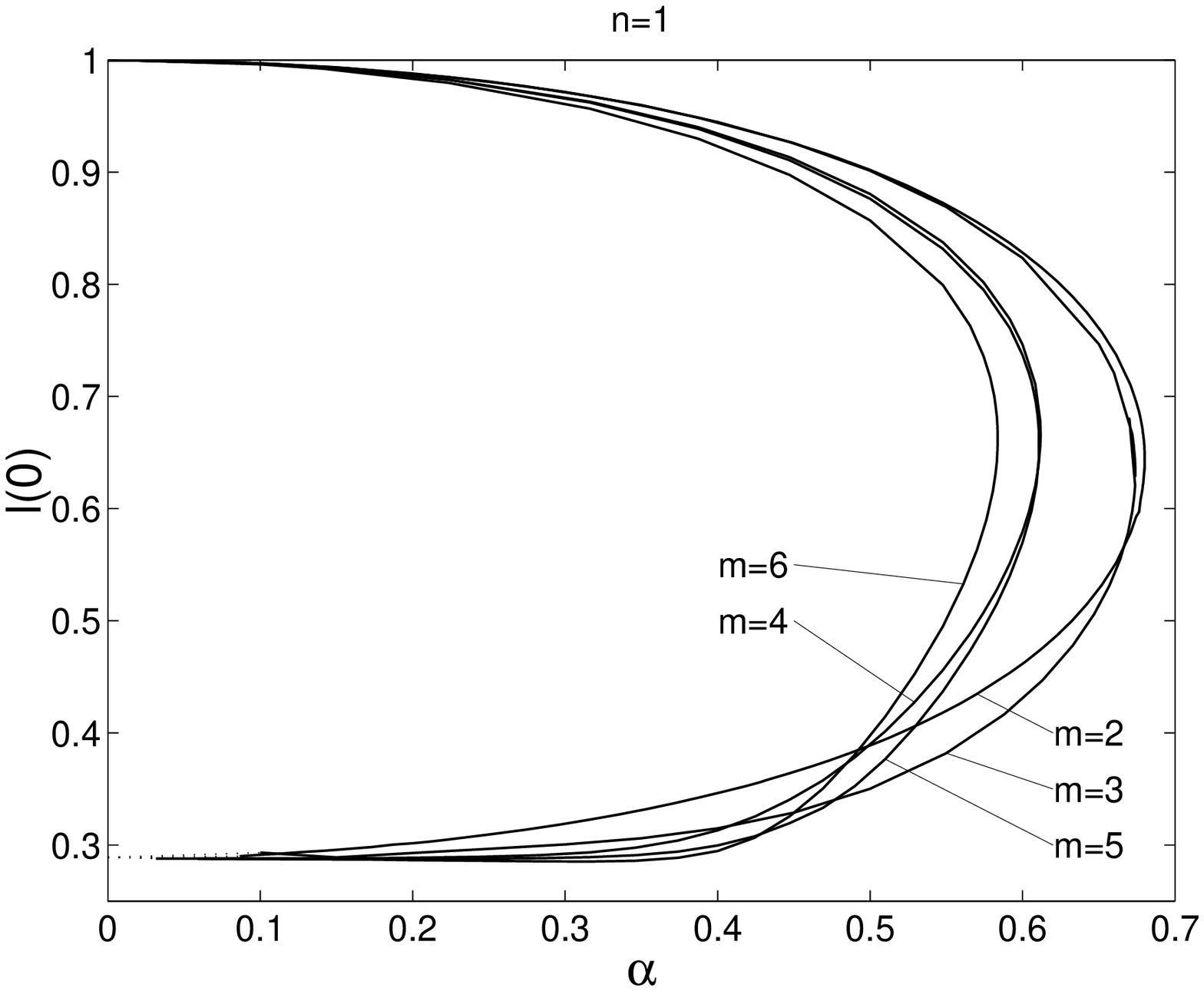} }
}\vspace{0.5cm}
{\bf Fig.~2} \small
The values of the metric functions $f$ (a) and $l$ (b) at the origin
are shown as functions of the coupling constant $\alpha$
for the chain solutions with $n=1$ and $m=2-6$.
The dotted lines extend the EYMH curves 
to values of the lowest mass Bartnik-McKinnon solution.
\vspace{0.5cm}
}

Let us now consider the nodes of the Higgs field for these even $m$ chains.
In Fig.~3a we recall the dependence of the single positive node of the 
monopole-antimonopole pair on the coupling constant.
Starting from the flat space limit, the location of the node continuously
moves inwards along both branches, and reaches the origin in the 
$\alpha \rightarrow 0$ limit on the upper branch.
(We do not consider excited monopole-antimonopole pair solutions with
several nodes here, 
which are related to excited Bartnik-McKinnon solutions \cite{mapKK}.)
The flat space $m=4$ solution has two positive nodes.
The evolution of these nodes along both branches is exhibited in
Fig.~3b.
The location of the inner node again moves continuously
inwards along both branches, and reaches the origin in the limit
$\alpha \rightarrow 0$ on the upper branch. In contrast, 
the location of the outer node reaches a finite limiting value 
in the limit $\alpha \rightarrow 0$ on the upper branch,
which, interestingly, agrees with the location of the single node of the
flat space monopole-antimonopole pair solution.
Similarly, as seen in Fig.~3c,
of the 3 positive nodes of the $m=6$ solution
the location of the innermost node reaches the origin in the limit
$\alpha \rightarrow 0$ on the upper branch, 
whereas the locations of the other two nodes 
reach finite limiting values,
which agree with the locations of the two nodes of the
flat space $m=4$ chain.

\noindent\parbox{\textwidth}{
\centerline{
(a)\mbox{\epsfysize=4.5cm \epsffile{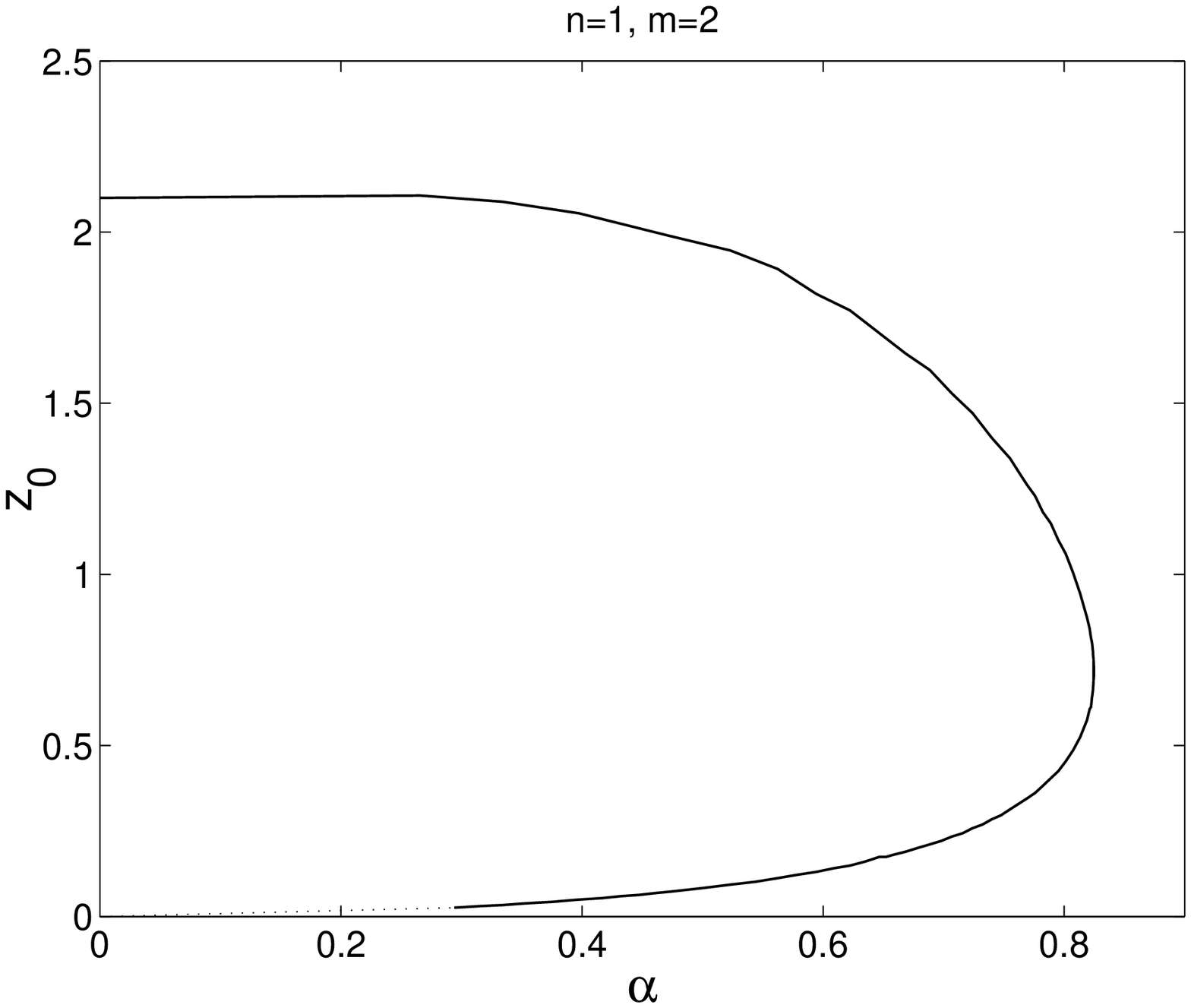} }
(b)\mbox{\epsfysize=4.5cm \epsffile{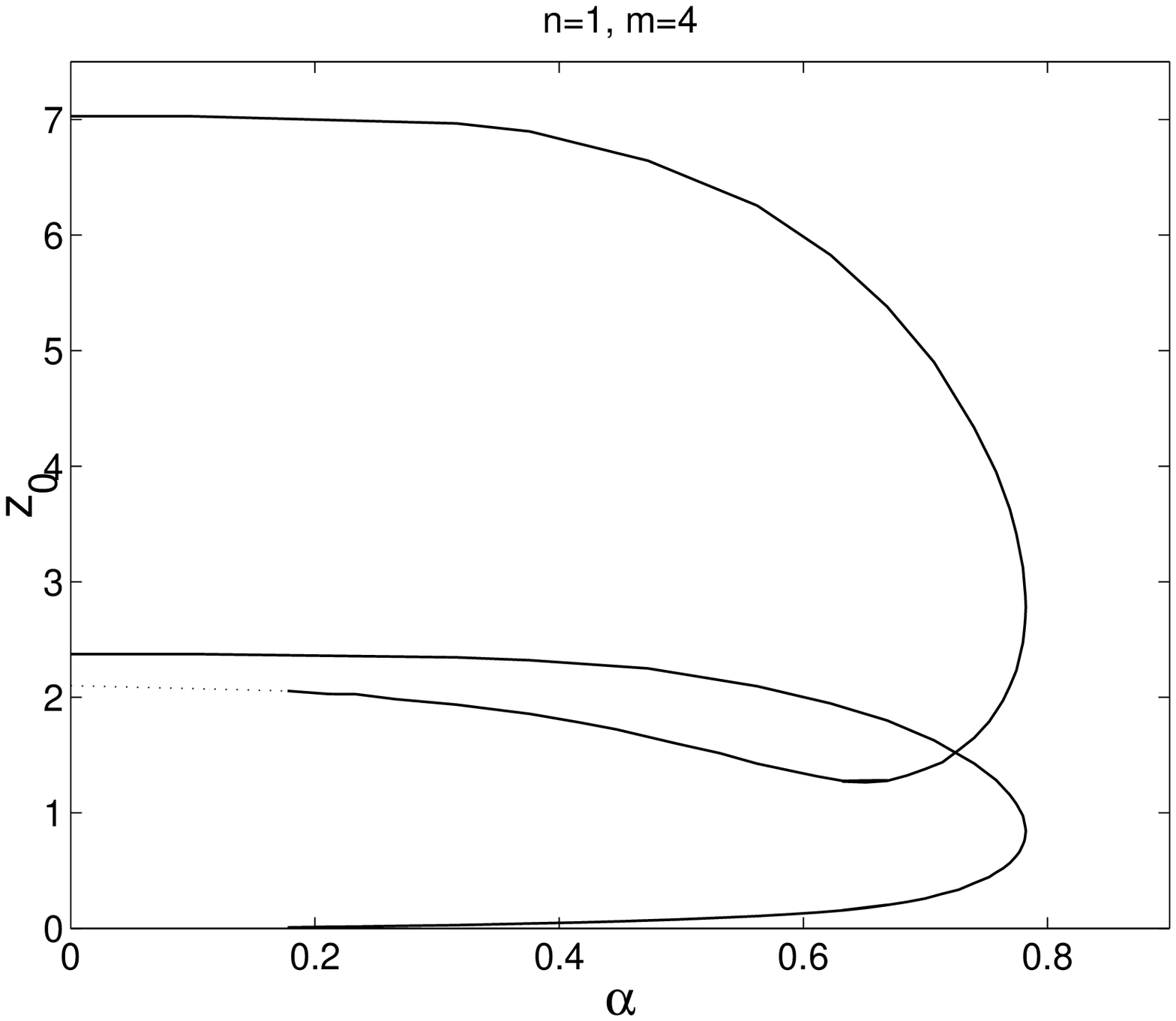} }
(c)\mbox{\epsfysize=4.5cm \epsffile{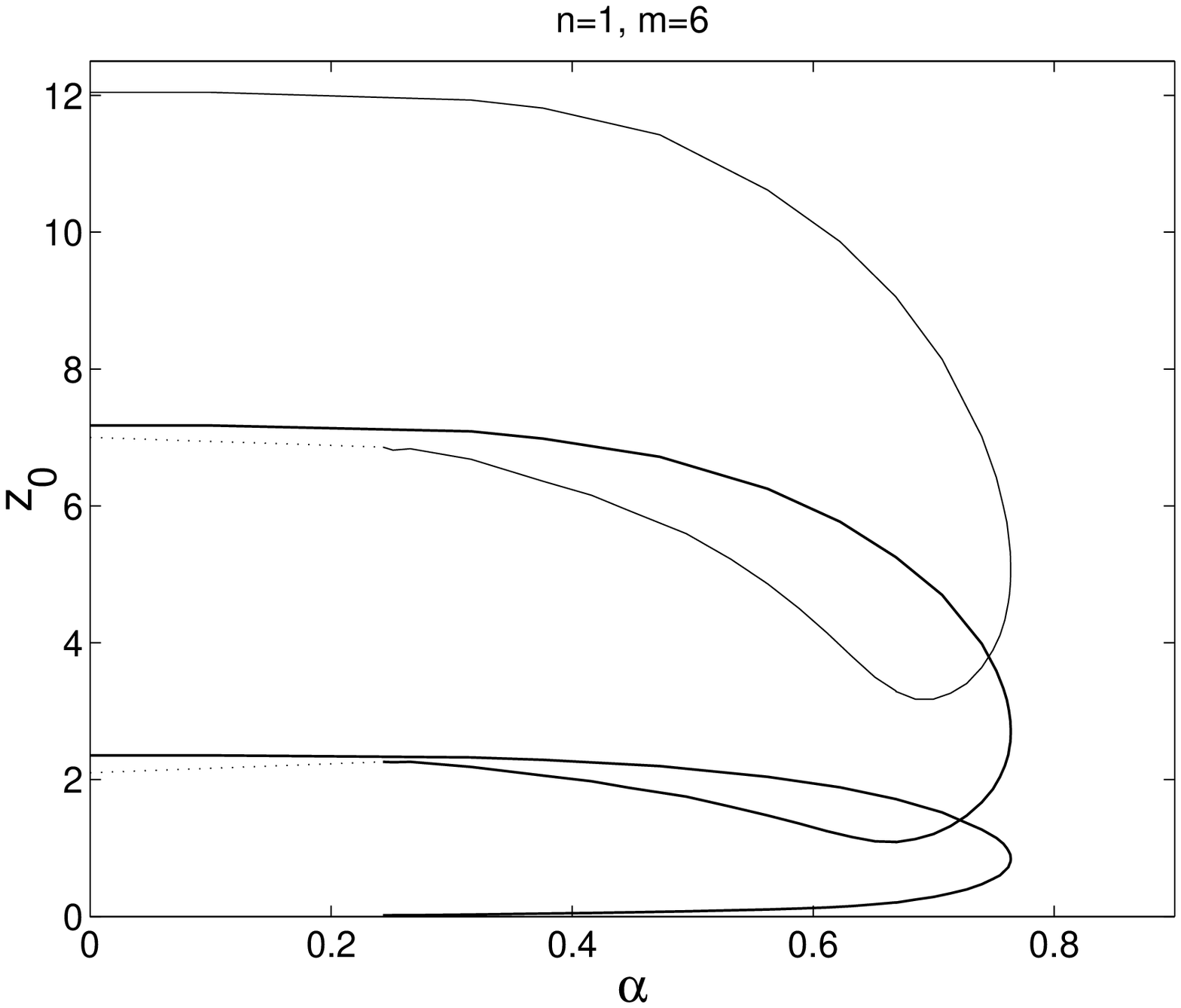} }
}\vspace{0.5cm}
\centerline{
(d)\mbox{\epsfysize=4.5cm \epsffile{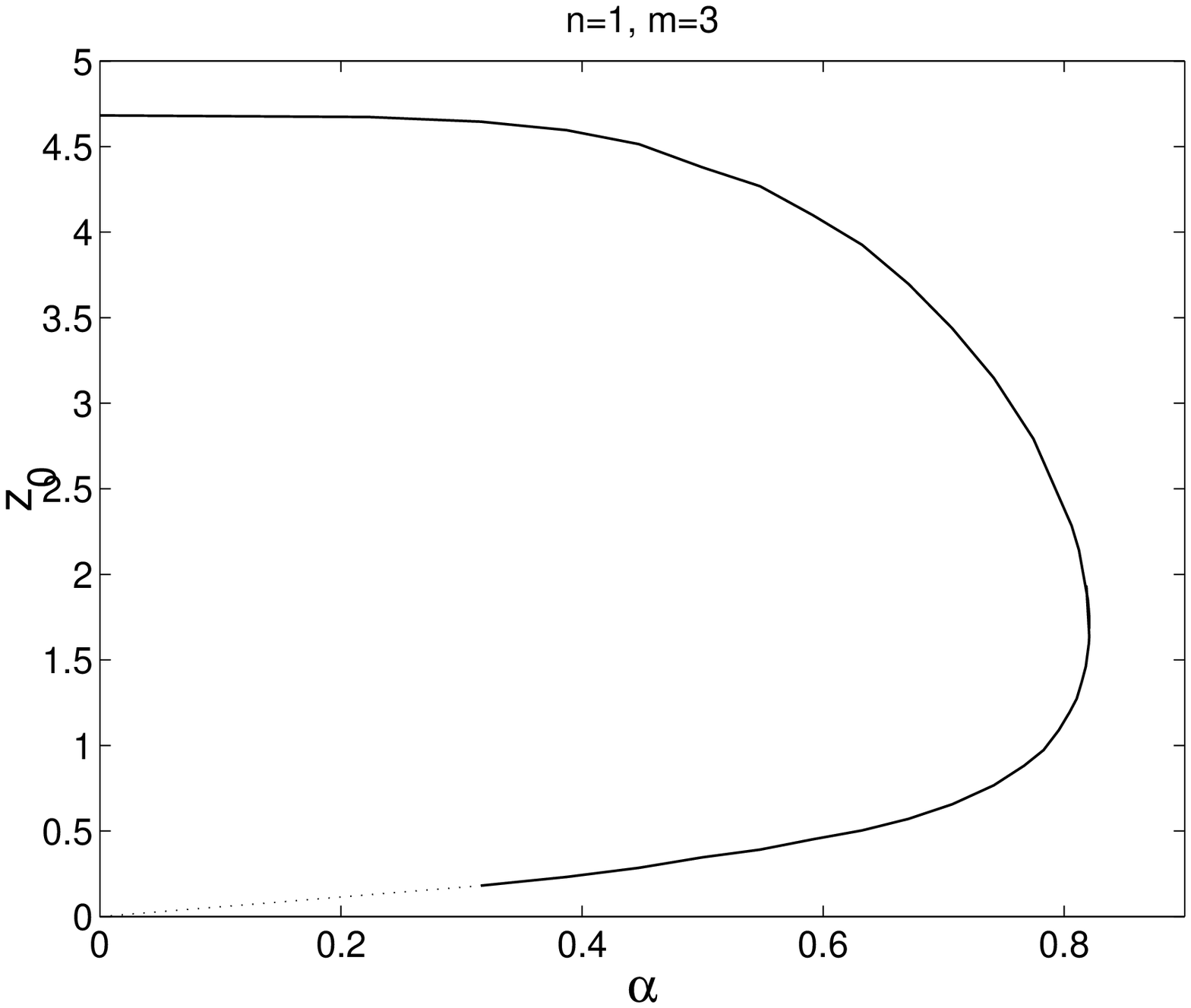} }
(e)\mbox{\epsfysize=4.5cm \epsffile{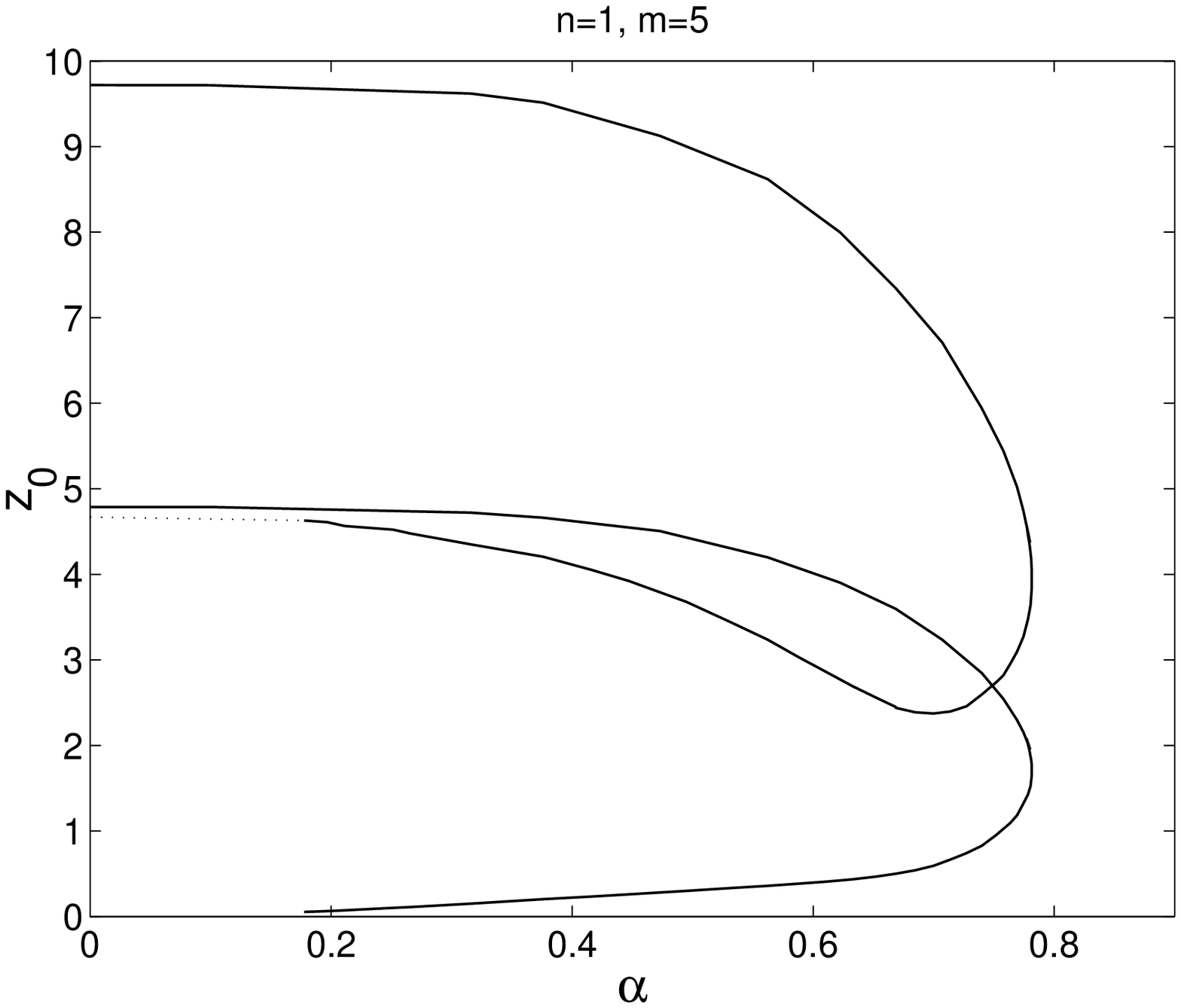} }
}\vspace{0.5cm}
{\bf Fig.~3} \small
The nodes of the Higgs field 
are shown as functions of the coupling constant $\alpha$
for the chain solutions with $n=1$ and $m=2$ (a), $m=4$ (b)
$m=6$ (c), $m=3$ (d), $m=5$ (e).
The dotted lines extend the nodes of the EYMH solutions
to zero or to the respective values of the $m-2$ YMH solutions.
\vspace{0.5cm}
}

To obtain a better understanding of the limit $\alpha \rightarrow 0$ 
on the upper branch for these chains, let us consider 
the matter and metric functions for small values of the coupling constant.
In Fig.~4, for instance, we exhibit the gauge field function $K_2$,
the modulus of the Higgs field $|\Phi|$ and the metric function $f$
for the $m=4$ chain
for a small value of the coupling constant, $\alpha^2=0.001$, 
on the upper branch.
We observe, that we have to distinguish two regions for
the matter function $K_2$, an outer region and an inner region.
In the outer region, the gauge field function
$K_2$ of the $m=4$ chain agrees well with the (by 2 shifted) corresponding
gauge field function $K_2$ of the 
flat space ($m=2$) monopole-antimonopole pair solution,
while in the inner region the gauge field function $K_2$ 
of the $m=4$ chain agrees well
with the corresponding function of the BM solution
(in by $\alpha$ scaled coordinates).
Likewise, in the outer region
the modulus of the Higgs field $|\Phi|$ of the $m=4$ chain
agrees well with the modulus of the Higgs field of the
flat space ($m=2$) monopole-antimonopole pair solution.
In the BM solution, of course, no Higgs field is present.
(An analogous pattern holds for the functions $K_1$, $K_3$, $K_4$,
$\Phi_1$ and $\Phi_2$ after a gauge transformation.)
The metric function $f$, on the other hand,
agrees well with the metric function of
the BM solution (in by $\alpha$ scaled coordinates) everywhere.
(The same holds for the functions $m$ and $l$.)
The metric of the flat space ($m=2$) monopole-antimonopole pair solution
is, of course, trivial.

Thus, for small $\alpha$ on the upper branch,
the $m=4$ chain may be thought of as composed of a scaled
BM solution in the inner region and a flat space 
($m=2$) monopole-antimonopole pair solution in the outer region.
Taking the limit $\alpha \rightarrow 0$ then, the inner region
shrinks to zero size, yielding a solution which is singular at the origin
and which has infinite mass, while elsewhere it corresponds to the
flat space ($m=2$) monopole-antimonopole pair solution.
This explains the above observation concerning the location of the Higgs
field nodes, made in Fig.~3.
On the other hand, taking the limit $\alpha \rightarrow 0$
with scaled coordinates, the inner region covers all of space
and contains the BM solution with its finite scaled mass,
while the flat space ($m=2$) monopole-antimonopole pair solution
is shifted all the way out to infinity.

\noindent\parbox{\textwidth}{
\centerline{
(a)\mbox{\epsfysize=6.0cm \epsffile{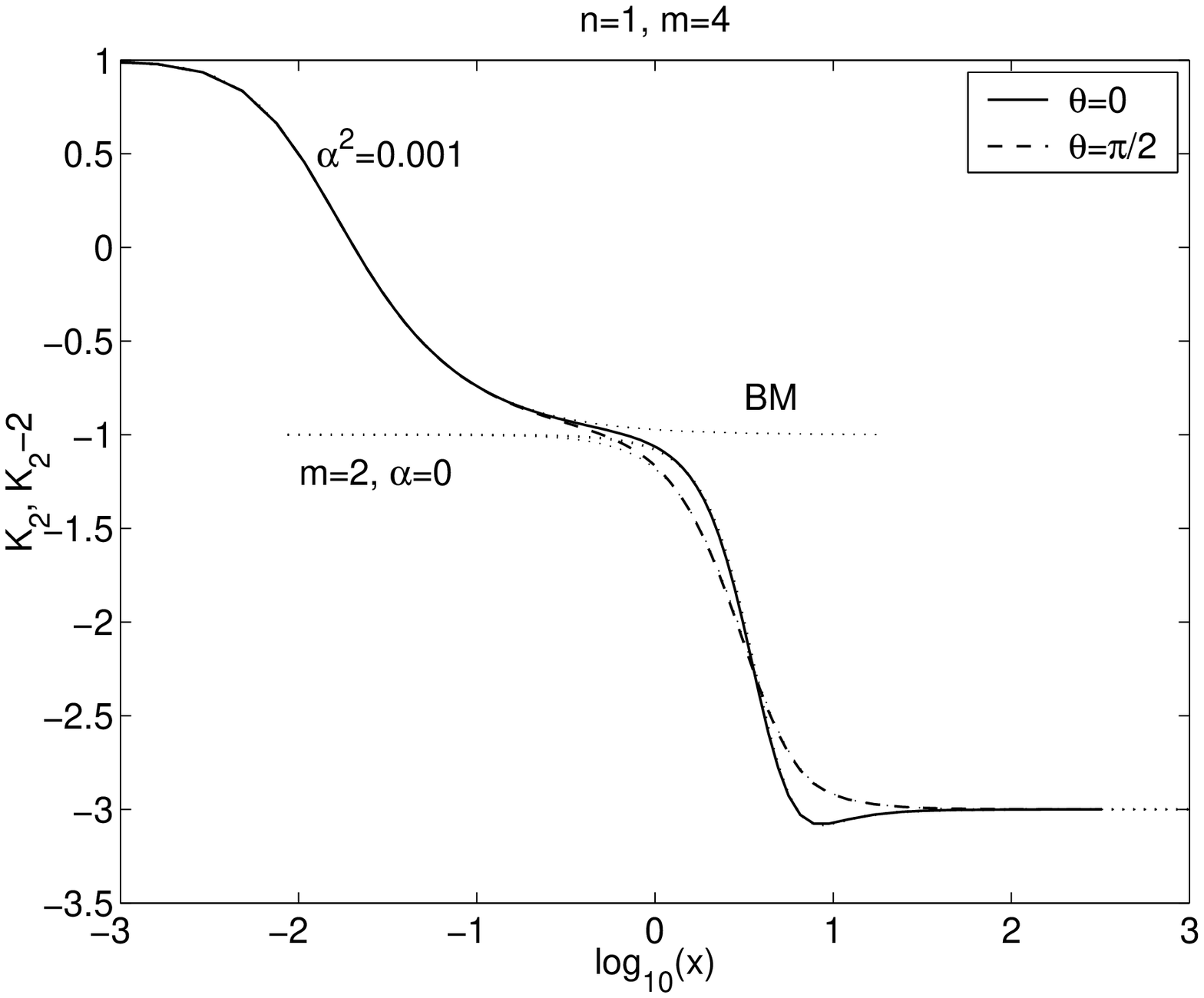} }
(b)\mbox{\epsfysize=6.0cm \epsffile{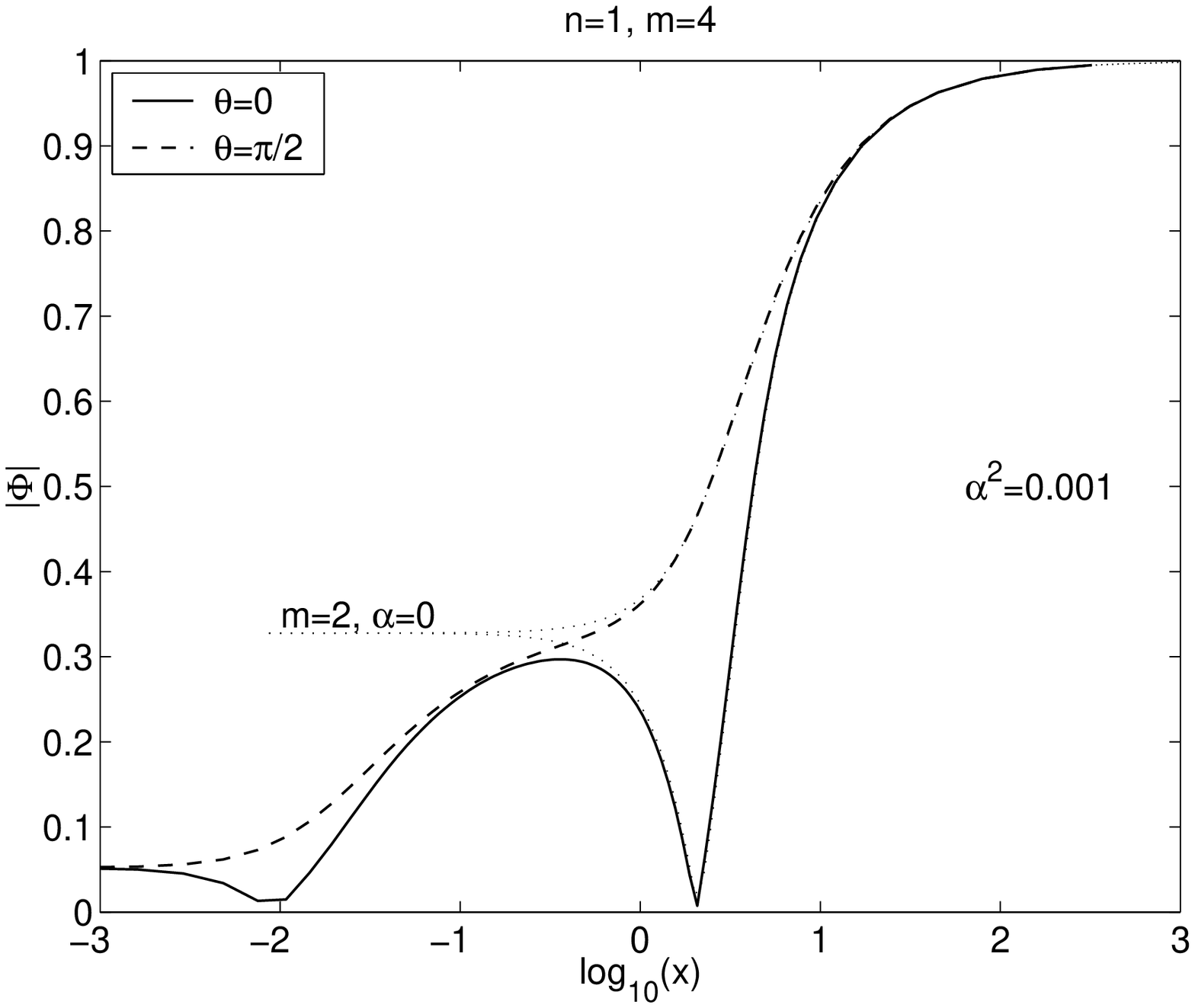} }
}\vspace{0.5cm}
\centerline{
(c)\mbox{\epsfysize=6.0cm \epsffile{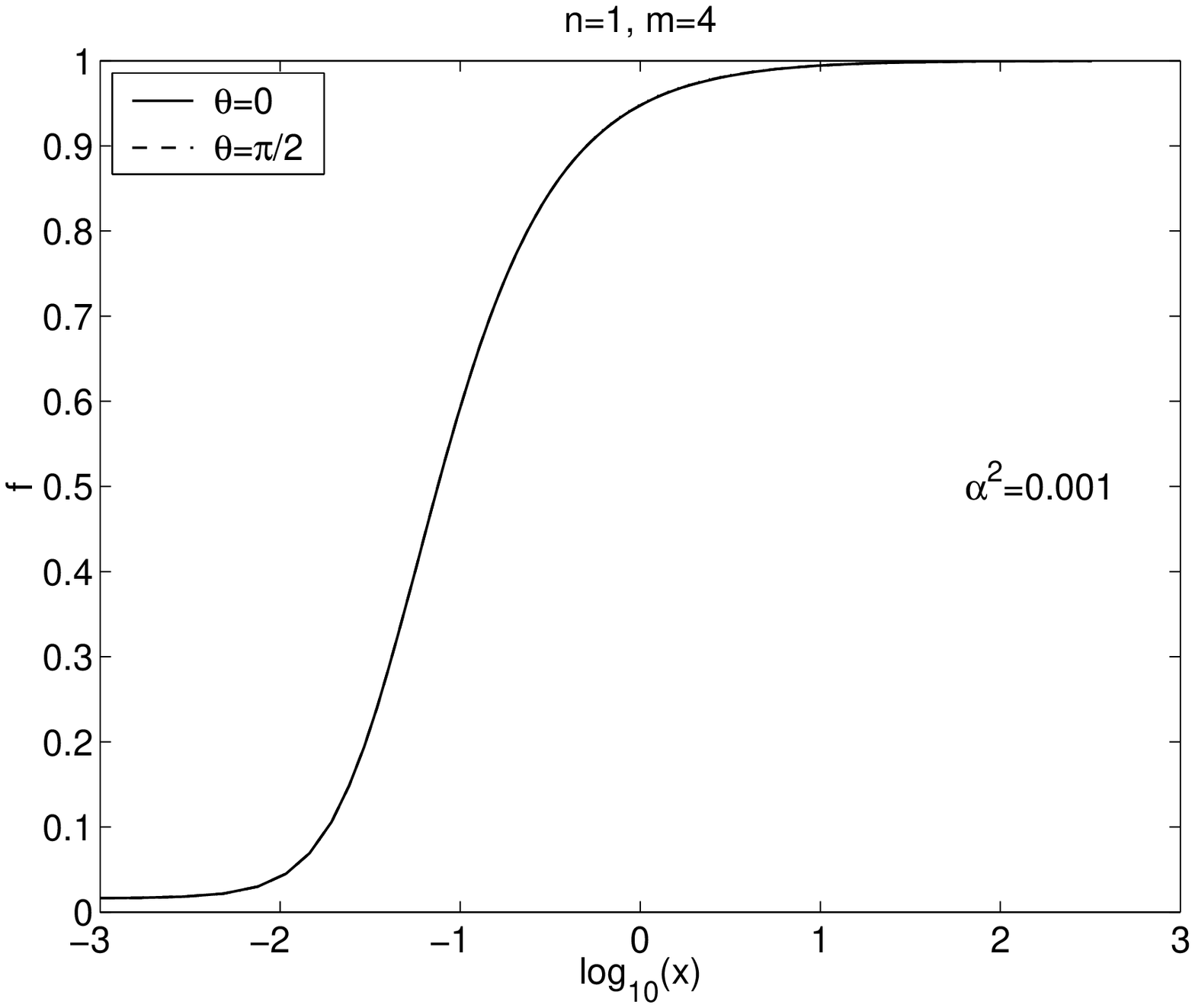} }
}\vspace{0.5cm}
{\bf Fig.~4} \small
The gauge field function $K_2$ (a), the modulus of the Higgs field 
$|\Phi|$ (b) and the metric function $f$ (c)
are shown for the chain solution with $n=1$ and $m=4$ at $\alpha^2=0.001$.
Also shown are the functions $K_2$ and $|\Phi|$ of the flat space
monopole-antimonopole solution, and the functions $K_2$ and $f$
of the lowest mass Bartnik-McKinnon solution (in scaled coordinates).
\vspace{0.5cm}
}

For the $m=6$ chain, we observe an analogous pattern for the functions
in the limit $\alpha \rightarrow 0$ on the upper branch,
as exhibited in Fig.~5.
In the outer region, the gauge field function
$K_2$ of the $m=6$ chain agrees well with the (by 2 shifted) corresponding
gauge field function $K_2$ of the flat space $m=4$ chain,
while in the inner region it agrees well with the corresponding BM function
(in by $\alpha$ scaled coordinates).
Likewise, in the outer region
the modulus of the Higgs field $|\Phi|$ of the $m=6$ chain
agrees well with the modulus of the Higgs field of the
flat space $m=4$ chain,
while the metric function $f$ of the $m=6$ chain
agrees well with the metric function of
the BM solution (in by $\alpha$ scaled coordinates) everywhere.

\noindent\parbox{\textwidth}{
\centerline{
(a)\mbox{\epsfysize=6.0cm \epsffile{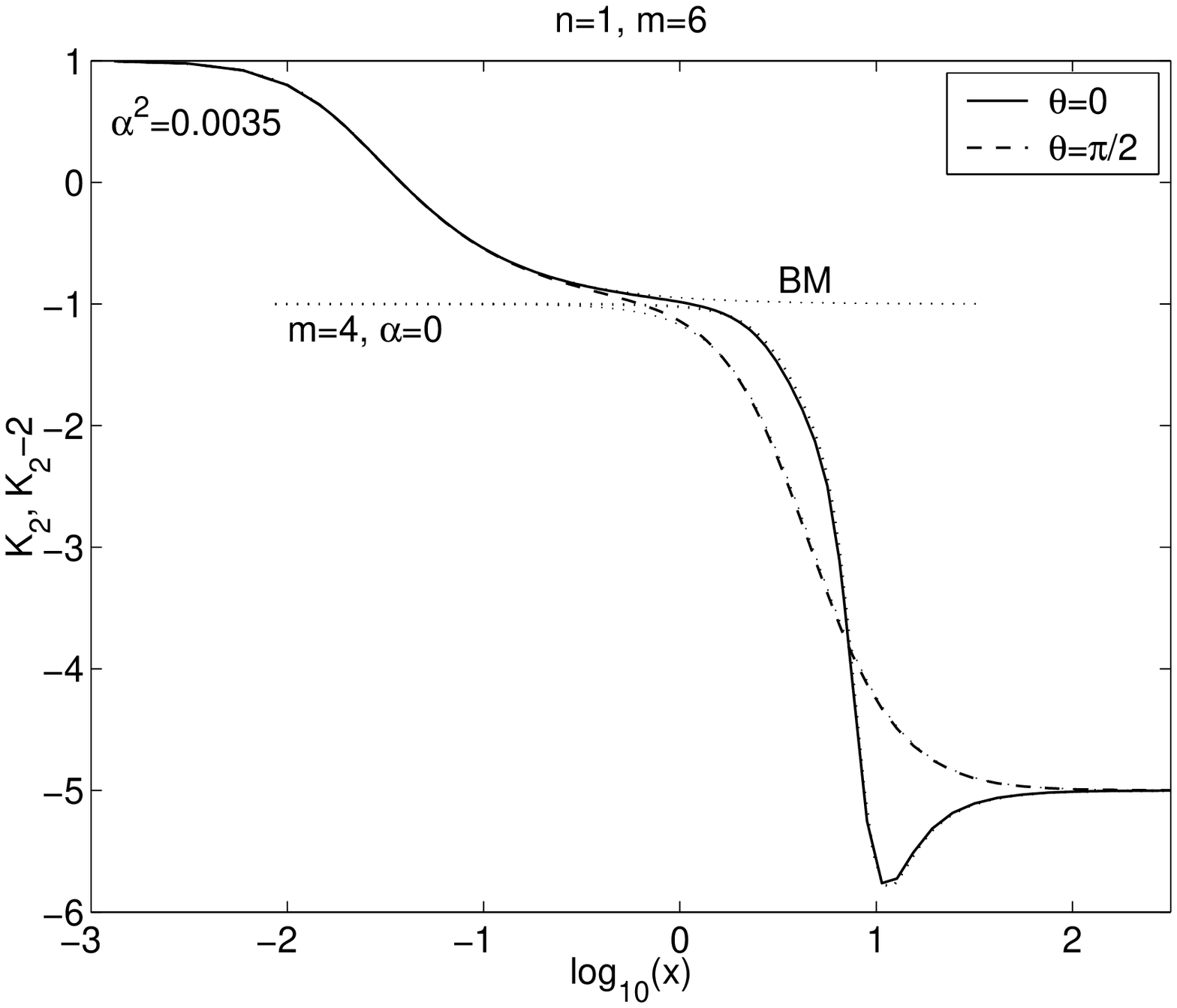} }
(b)\mbox{\epsfysize=6.0cm \epsffile{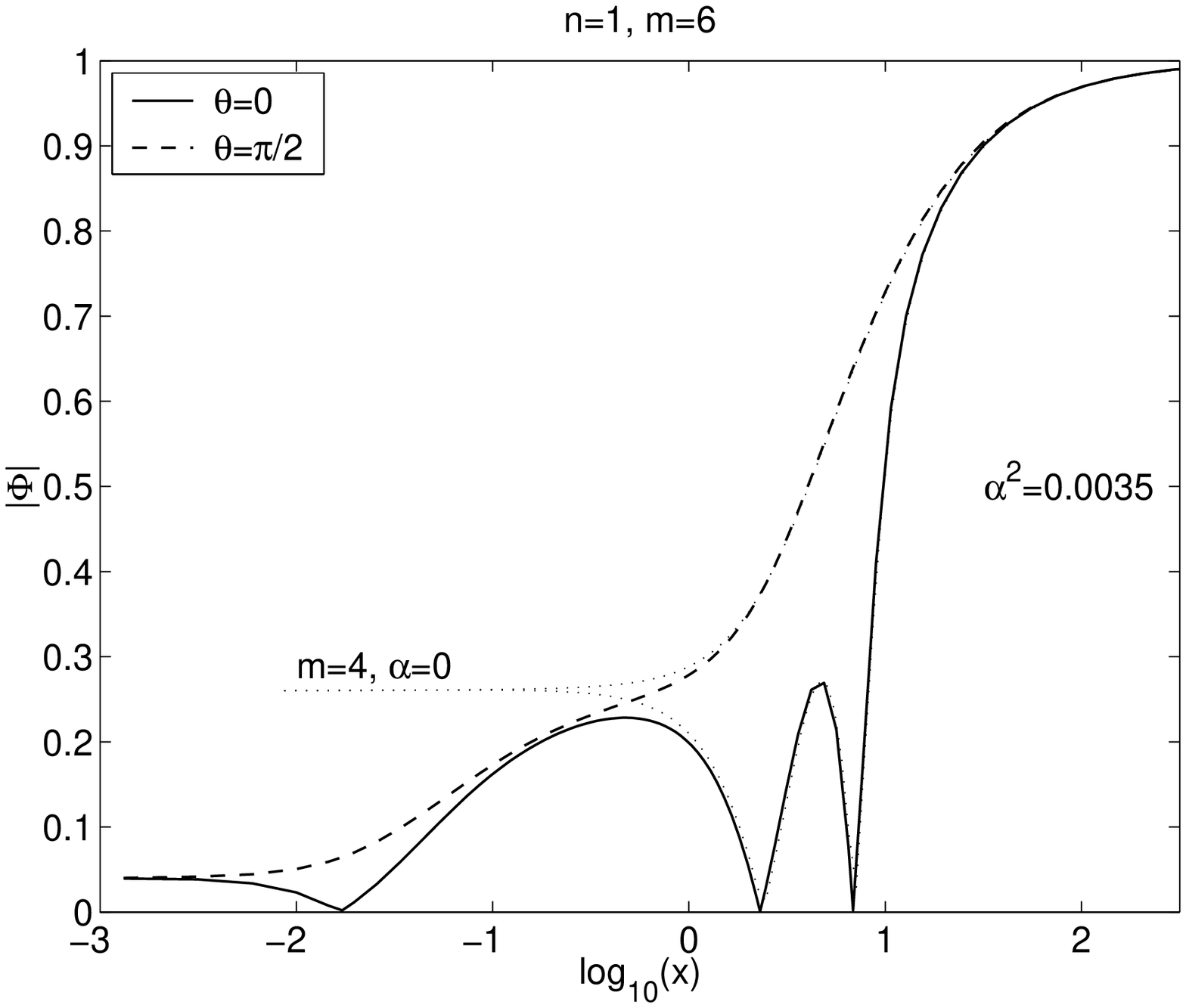} }
}\vspace{0.5cm}
{\bf Fig.~5} \small
The gauge field function $K_2$ (a), the modulus of the Higgs field 
$|\Phi|$ (b) 
are shown for the chain solution with $n=1$ and $m=6$ at $\alpha^2=0.0035$.
Also shown are the functions $K_2$ and $|\Phi|$ of the flat space
$m=4$ chain, and the functions $K_2$ and $f$
of the lowest mass Bartnik-McKinnon solution (in scaled coordinates).
\vspace{0.5cm}
}

Turning now to monopole-antimonopole chains in the
topological sector with unit charge,
one might expect a completely different pattern for
the coupling constant dependence of the solutions, namely
a pattern which would be similar to the coupling constant 
dependence of the gravitating monopoles \cite{gmono},
where the upper branch of gravitating monopoles bifurcates
with the branch of extremal RN solutions \cite{gmono,foot}.
Such an expectation does not prove true, however.
Instead, the odd $m$ chains show a similar pattern as
the even $m$ chains.
Two branches of gravitating $m$ chain solutions merge at
a maximal value $\alpha_{\rm max}$. The lower branch of solutions
emerges from the flat space $m$ chain,
and the upper branch extends all the way back to $\alpha=0$. 
For small values of $\alpha$ on the upper branch,
the solutions may be thought of as composed of a scaled
BM solution in the inner region and a flat space
$m-2$ solution in the outer region.

For the $m=3$ and $m=5$ chains,
we exhibit the mass and scaled mass in Fig.~1, 
the values of the metric functions $f$ and $l$
at the origin in Fig.~2, and the location of the positive node(s),
in Fig.~3. Note, that 
in the limit $\alpha \rightarrow 0$ on the upper branch,
the location of the outer positive node of the $m=5$ chain
agrees with the location of the positive node of the
flat space $m=3$ chain.
The composition of the solutions in terms of a scaled
BM solution in the inner region and a flat space
$m-2$ solution in the outer region is illustrated in Fig.~6,
where we exihibit the gauge field function $K_2$
for the $m=3$ and $m=5$ chains.
For the $m=3$ chain clearly the gauge field function corresponds 
in the outer region to the function of a flat space ($m=1$) monopole,
while for the $m=5$ chain it corresponds to a flat space
$m=3$ chain.

\noindent\parbox{\textwidth}{
\centerline{
(a)\mbox{\epsfysize=6.0cm \epsffile{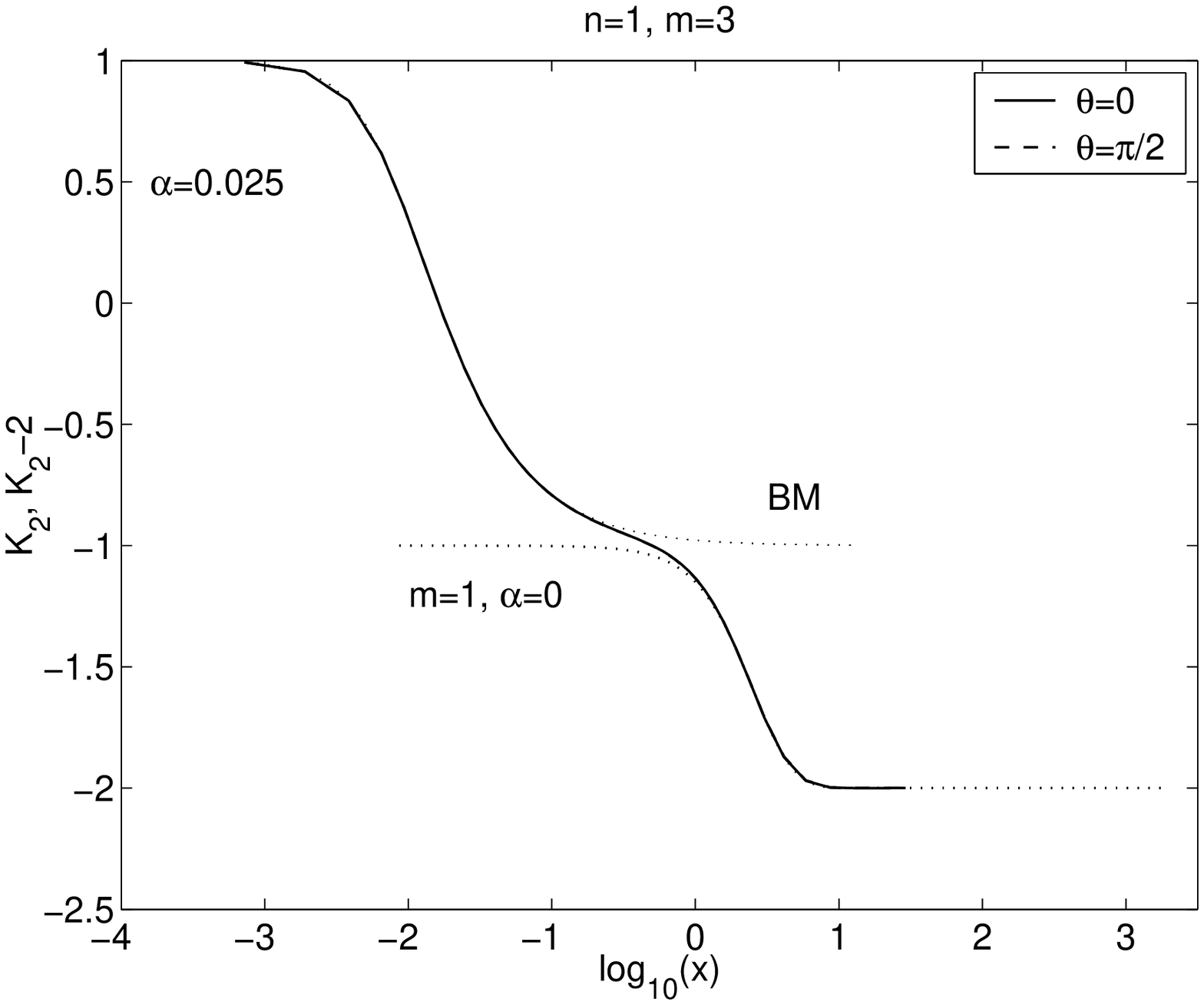} }
(b)\mbox{\epsfysize=6.0cm \epsffile{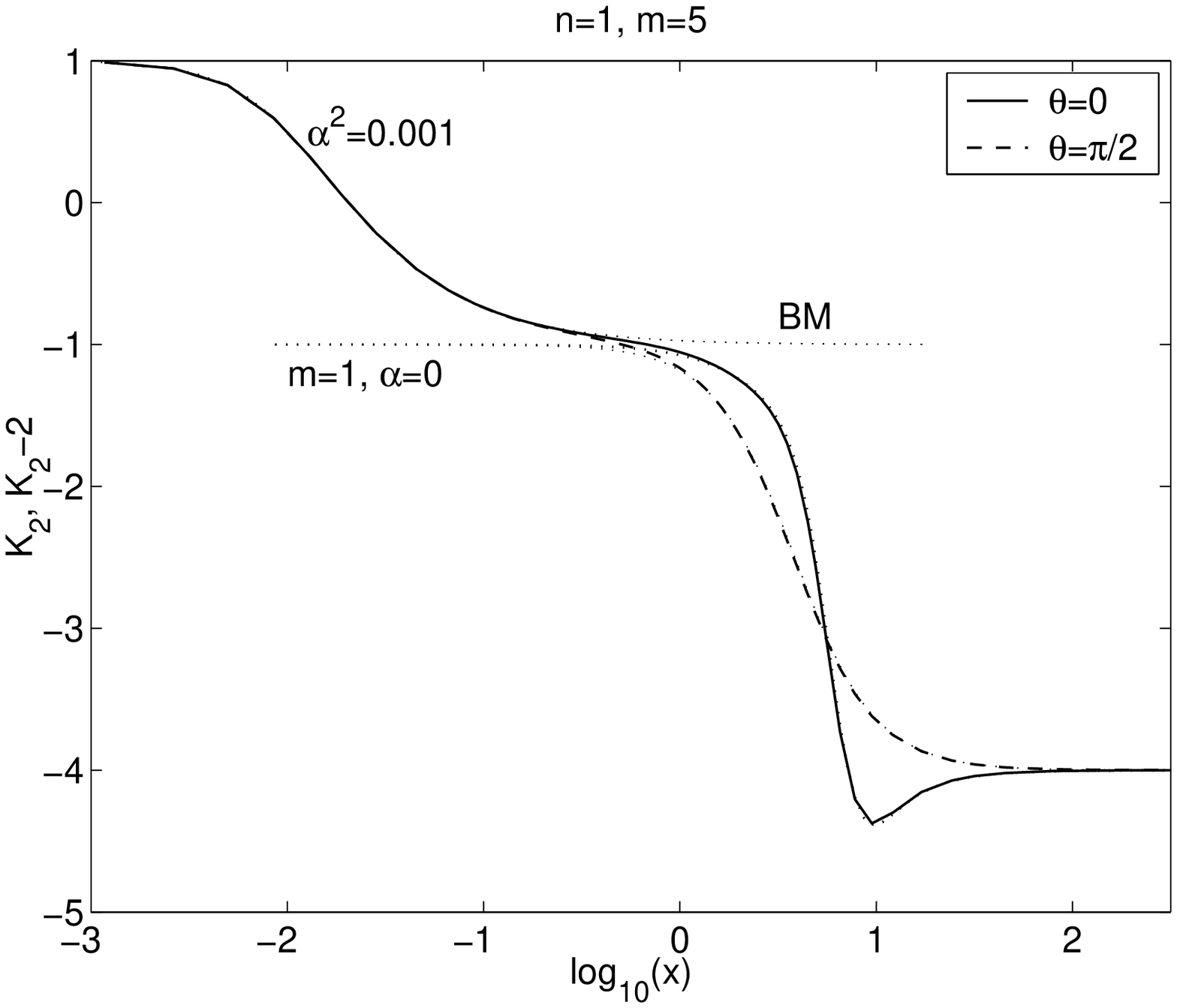} }
}\vspace{0.5cm}
{\bf Fig.~6} \small
The gauge field function $K_2$ is shown for the chain solutions 
with $n=1$ and $m=3$ at $\alpha=0.025$ (a) and
with $n=1$ and $m=5$ at $\alpha^2=0.001$ (b).
Also shown are the functions $K_2$ of the flat space
monopole and $m=3$ solutions, and the function $K_2$
of the lowest mass Bartnik-McKinnon solution (in scaled coordinates).
\vspace{0.5cm}
}

Concluding, we observe the same general pattern for all gravitating
chain solutions with $n=1$.
From the flat space $m$ chain a lower branch of gravitating $m$ chains
emerges, which merges at a maximal value $\alpha_{\rm max}$
with an upper branch. 
The value of $\alpha_{\rm max}$ decreases with increasing $m$.
For small $\alpha$ the solutions on the upper branch
may be thought of as composed of a scaled
BM solution in the inner region and a flat space
$m-2$ solution in the outer region.
Consequently, the mass diverges in the limit $\alpha \rightarrow 0$,
while the scaled mass approaches the mass of the lowest BM solution
of EYM theory.

\boldmath
\subsection{Gravitating chains: $n=2$}
\unboldmath

For the $m$ chains composed of monopoles and antimonopoles with
charge $n=2$ \cite{KKS,tigran}, 
we observe a completely analogous pattern
as for the chains composed of monopoles and antimonopoles with
charge $n=1$.
For small $\alpha$ the solutions on the upper branch
may be thought of as composed of a scaled generalized BM
solution with $n=2$ in the inner region \cite{KK},
and a flat space $m-2$ solution in the outer region.
Consequently, the mass again diverges in the limit $\alpha \rightarrow 0$,
while the scaled mass approaches the mass of the 
generalized BM solution with $n=2$.

We illustrate this behaviour in Fig.~7 for the $m=4$ chain with $n=2$.
Clearly, the upper branch of the scaled mass 
approaches the mass of the generalized BM solution with $n=2$,
and the values of the metric functions $f$ and $l$ at the origin
approach those of the generalized BM solution with $n=2$ as well.
The composition of the solutions on the upper branch
for small values of $\alpha$ in terms of an inner 
generalized BM solution with $n=2$
and an outer flat space $m=2$ solution
is exhibited in Fig.~8 for the gauge field function $K_2$,
the modulus of the Higgs field $\Phi$ and the metric function $f$.

\noindent\parbox{\textwidth}{
\centerline{
(a)\mbox{\epsfysize=6.0cm \epsffile{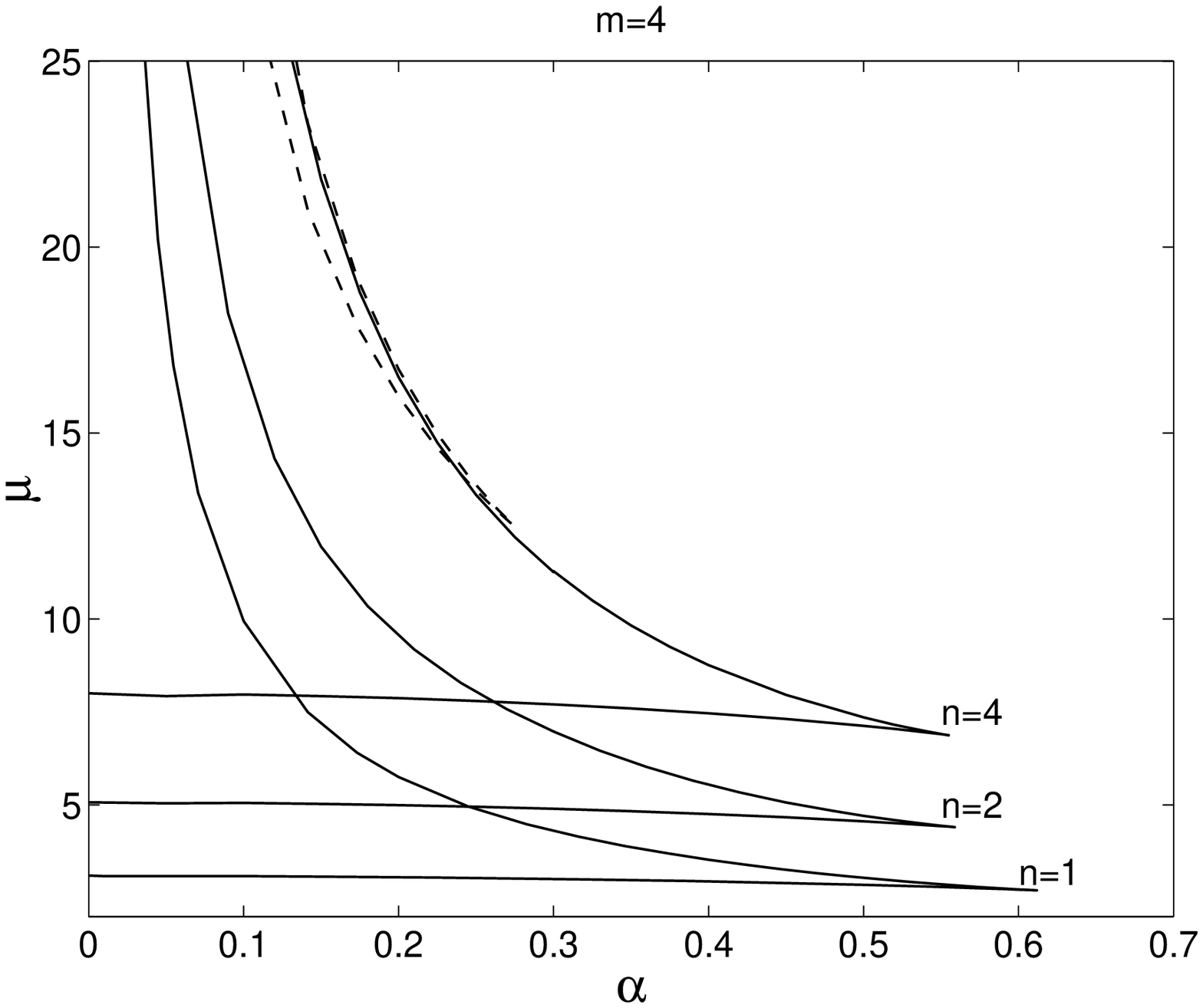} }
(b)\mbox{\epsfysize=6.0cm \epsffile{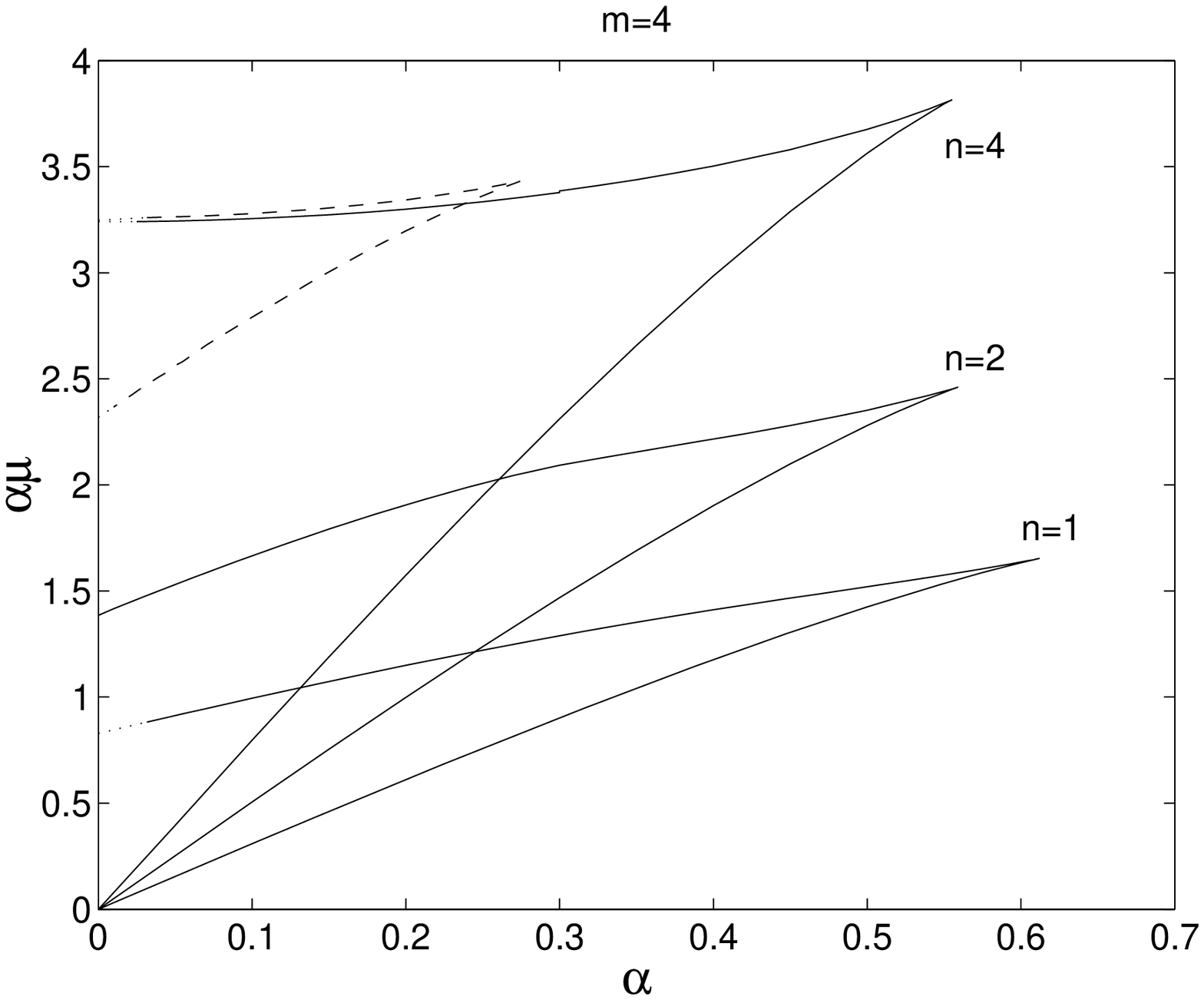} }
}\vspace{0.5cm}
{\bf Fig.~7} \small
The mass $\mu$ (a) and the scaled mass $\hat \mu$ (b)
are shown as functions of the coupling constant $\alpha$
for the chain solutions with $n=1$, 2, 4 and $m=4$.
The dotted lines extend the EYMH curves of the scaled mass to 
the masses of the lowest (generalized) Bartnik-McKinnon solutions.
\vspace{0.5cm}
}

\noindent\parbox{\textwidth}{
\centerline{
(a)\mbox{\epsfysize=6.0cm \epsffile{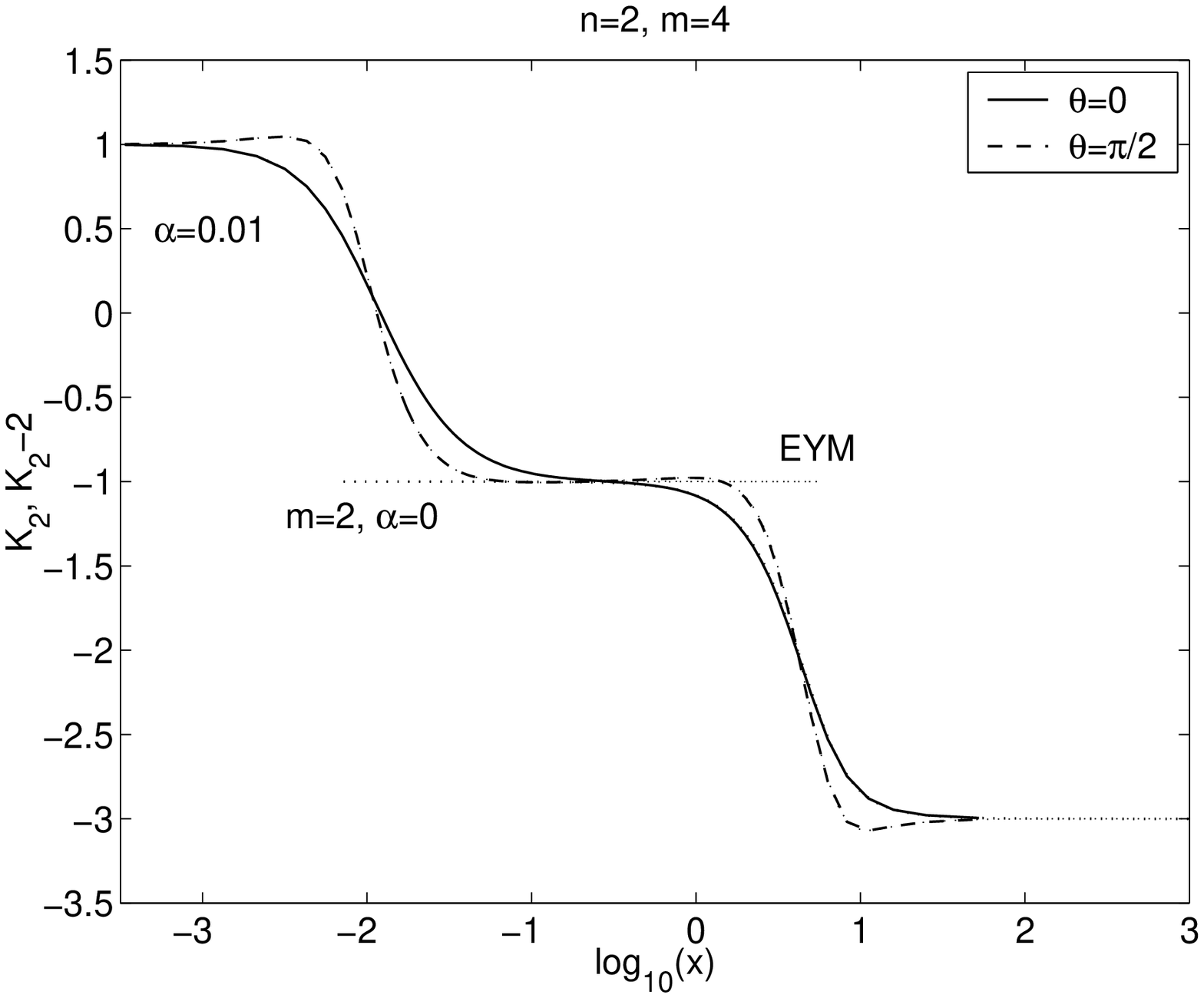} }
(b)\mbox{\epsfysize=6.0cm \epsffile{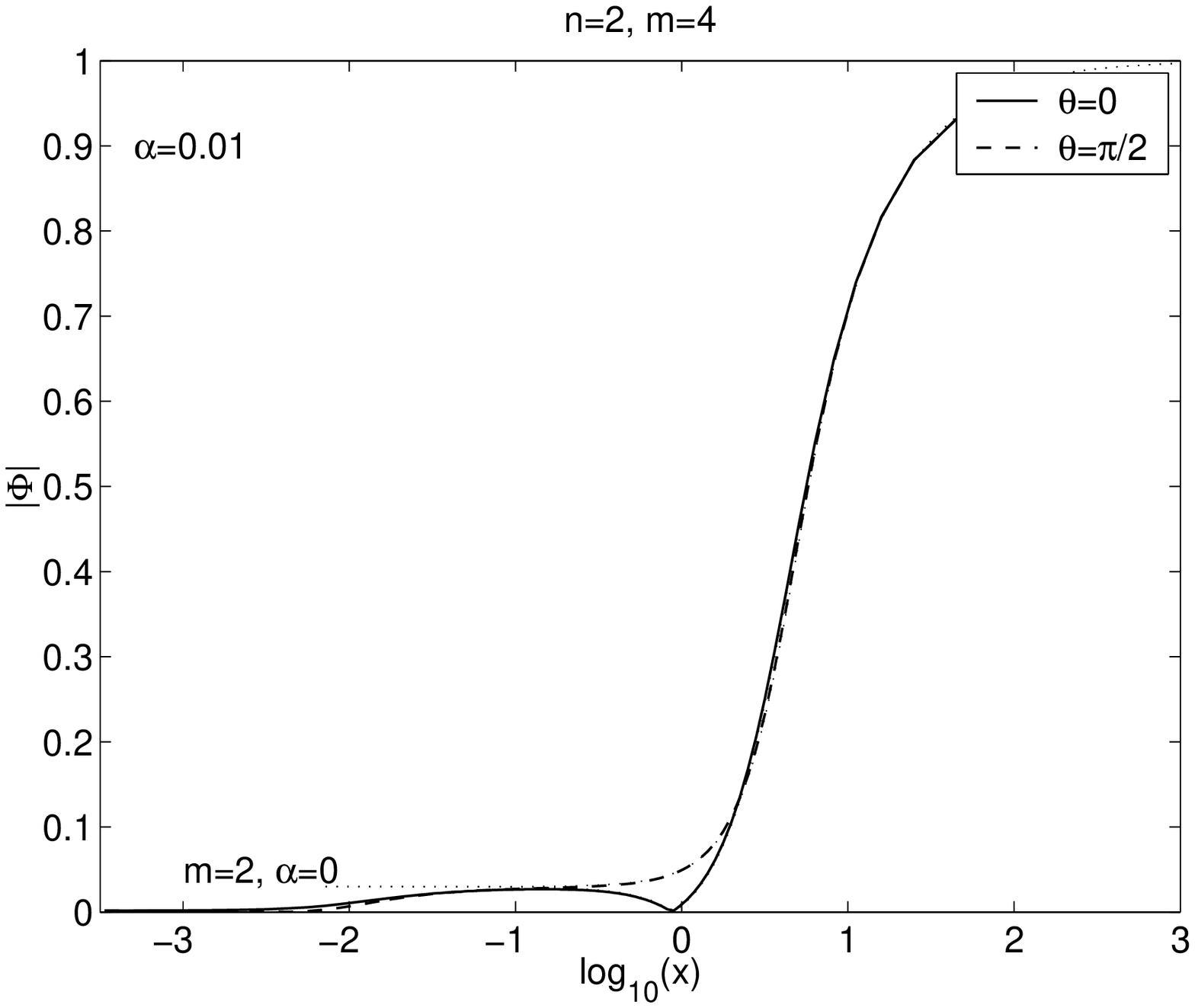} }
}\vspace{0.5cm}
\centerline{
(c)\mbox{\epsfysize=6.0cm \epsffile{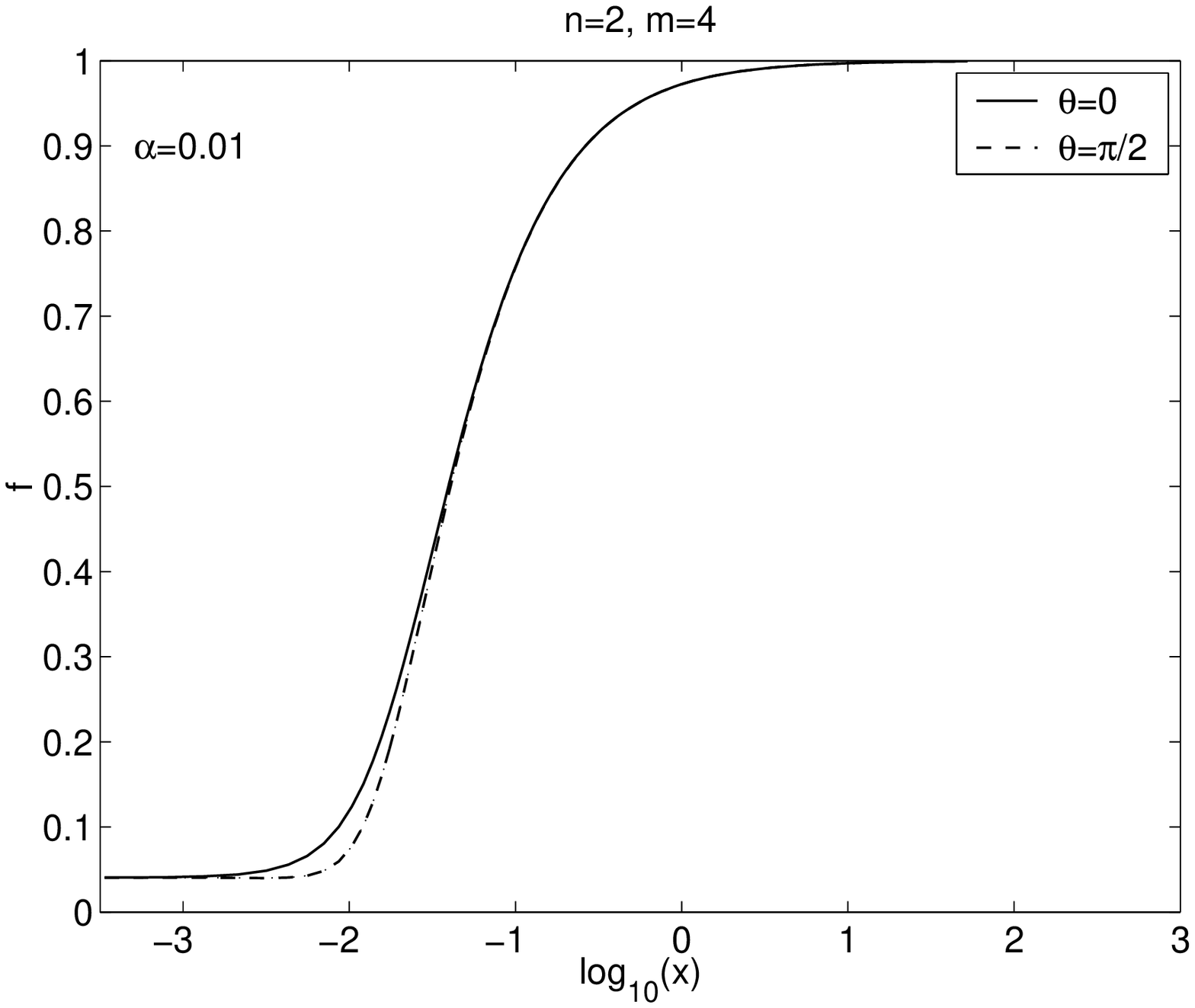} }
}\vspace{0.5cm}
{\bf Fig.~8} \small
The gauge field function $K_2$ (a), the modulus of the Higgs field 
$|\Phi|$ (b) and the metric function $f$ (c)
are shown for the chain solution with $n=2$ and $m=4$ at $\alpha=0.01$.
Also shown are the functions $K_2$ and $|\Phi|$ of the flat space
$m=2$ chain, and the functions $K_2$ and $f$
of the generalized BM solution with $n=2$
(in scaled coordinates).
\vspace{0.5cm}
}

\boldmath
\subsection{Gravitating vortex rings: $n=3$}
\unboldmath

For larger values of $n$ the flat space solutions completely change character,
since the structure of the nodes of the Higgs field 
in these solutions is totally different \cite{KKS}.
Whereas $m$ chains (with $n \le 2$)
possess only isolated nodes on the symmetry axis,
for the solutions with $n \ge 3$ the modulus of the Higgs field 
vanishes on rings in the $xy$-plane
or in planes parallel to the $xy$-plane,
centered around the $z$ axis. 
For even $m$ the solutions possess
only such vortex rings and no nodes on the symmetry axis.
For odd $m$ they possess vortex rings as well as a node at the origin,
where a charge $n$-monopole is located.

Let us now consider the effect of gravity on vortex solutions
with $n=3$.
Here the same general pattern is seen for the coupling constant dependence
as for the gravitating chains.
From the flat space vortex solution
a lower branch of gravitating vortex solutions
emerges, and merges at a maximal value $\alpha_{\rm max}$
with an upper branch.
On the upper branch, for small values of $\alpha$ the solutions
may be thought of as composed of a scaled
generalized BM solution with $n=3$ in the inner region
and a flat space $m-2$ solution in the outer region.
Thus the mass diverges in the limit $\alpha \rightarrow 0$ on the upper branch,
while the scaled mass approaches the mass of the 
generalized BM solution with $n=3$.

\boldmath
\subsection{Gravitating vortex rings: $n=4$}
\unboldmath

It is now tempting to conclude, that for arbitrary values of $m$ and $n$
the same general pattern holds:
From the flat space YMH solution
a lower branch of gravitating EYMH solutions
emerges, which merges at a maximal value $\alpha_{\rm max}$
with an upper branch.
On the upper branch, for small values of $\alpha$ the solutions
may be thought of as composed of a scaled 
generalized BM solution in the inner region
and a flat space $m-2$ solution in the outer region.
This conjectured pattern is indeed observed
for the $m=2$ and $m=3$ EYMH solutions with $n=4$.

For $n=4$ and $m=4$, however, a surprise in encountered.
As illustrated in Fig.~7, on the upper branch
the scaled mass of the $n=4$ and $m=4$ solutions 
does not approach the mass of
the generalized BM solution with $n=4$ \cite{KK},
but instead it approaches the mass of a new axially symmetric 
$n=4$ EYM solution \cite{IKKS}.
Moreover, a second new axially symmetric $n=4$ EYM solution exists,
which is slightly higher in mass \cite{IKKS}.
This second EYM solution constitutes the endpoint of a second
upper branch of $n=4$ and $m=4$ EYMH solutions, which merges at
a second maximal value of the coupling constant $\alpha$ with a
second lower branch, 
and it is along this second lower branch, 
that for small values of $\alpha$ the solutions
may be thought of as composed of a scaled
generalized BM solution with $n=4$ in the inner region
and a flat space $m=2$ solution in the outer region.
Thus for $n=4$ and $m=4$ there are four branches of solutions
instead of two, and consequently there are four limiting solutions 
when $\alpha \rightarrow 0$.
The masses of the generalized BM solutions and of the new
EYM solutions are exhibited in Fig.~9 \cite{IKKS}.
Note, that the boundary conditions of the new EYM solutions
differ from those of the generalized BM solutions at infinity \cite{IKKS}.
The EYM solutions can thus be classified by these boundary conditions.
Denoting $m=2k$, the new solutions have $k=2$, whereas the
generalized BM solutions have $k=1$ \cite{IKKS}.

\noindent\parbox{\textwidth}{
\centerline{
\mbox{\epsfysize=6.0cm \epsffile{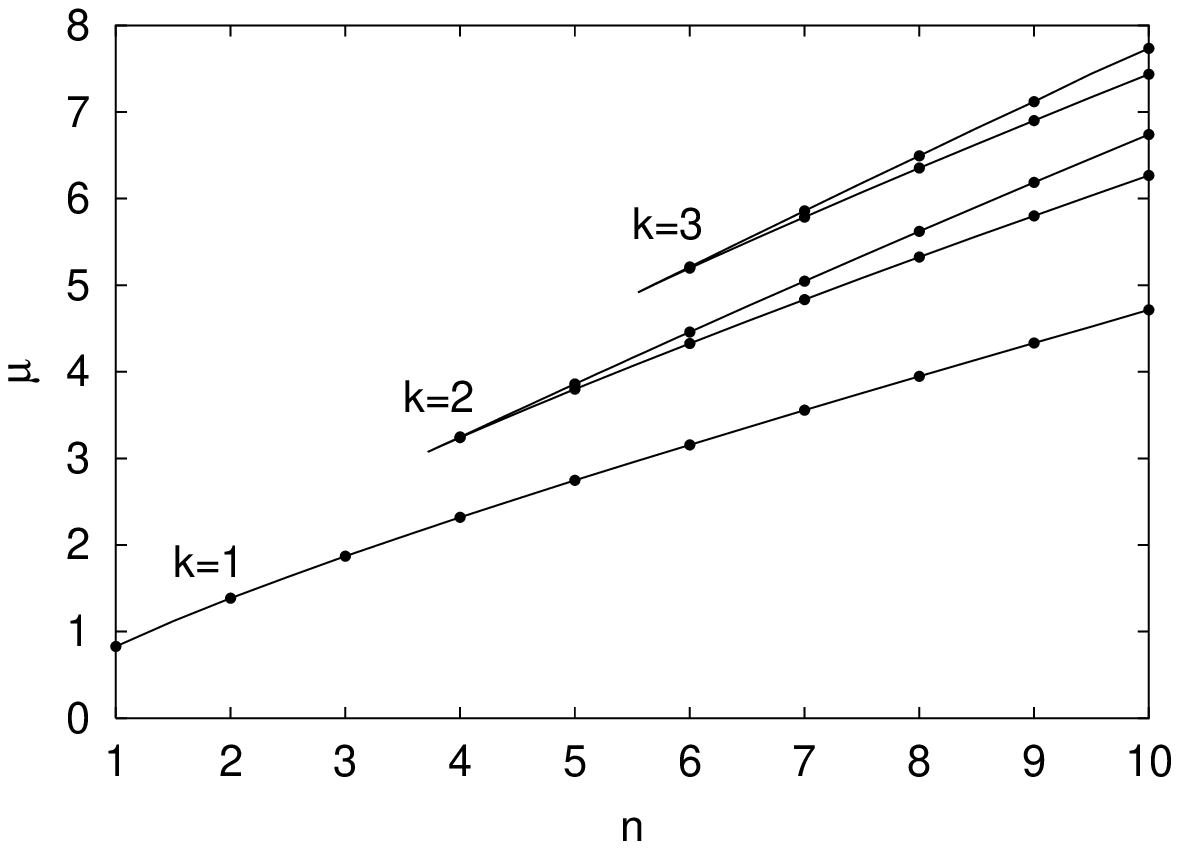} }
}\vspace{0.5cm}
{\bf Fig.~9} \small
The dependence of the mass $\mu$ on $n$ is shown
for the generalized BM solutions ($k=1$) and the new 
EYM solutions ($k=2$, 3).
\vspace{0.5cm}
}

In Fig.~10 we illustrate the dependence of the location
of the vortex of the Higgs field
on the coupling constant for these four branches of $n=4$ and $m=4$ solutions.
The flat space solution has two vortex rings,
located symmetrically in planes parallel to the $xy$ plane.
Starting from the flat space location,
the vortex rings move continuously inwards and towards the
$xy$ plane along the first lower branch and the first upper branch,
until at a critical
value of the coupling constant the two rings merge in the $xy$ plane.
When $\alpha$ is decreased further along the first upper branch,
a bifurcation takes place and two distinct rings appear in the
$xy$ plane.
In the limit $\alpha \rightarrow 0$ they shrink to zero size.
In scaled coordinates, in contrast, they reach a finite limiting size.
Likewise, in scaled coordinates there are two distinct rings
with finite size on the second upper branch (located not too far
from the two rings of the first upper branch).
Following the second upper branch and then the second lower branch,
the inner ring shrinks to zero size in the limit $\alpha \rightarrow 0$
(in ordinary coordinates), while the outer ring approaches
the location of the flat space vortex ring of the $n=4$ and $m=2$ solution.

\noindent\parbox{\textwidth}{
\centerline{
(a)\mbox{\epsfysize=6.0cm \epsffile{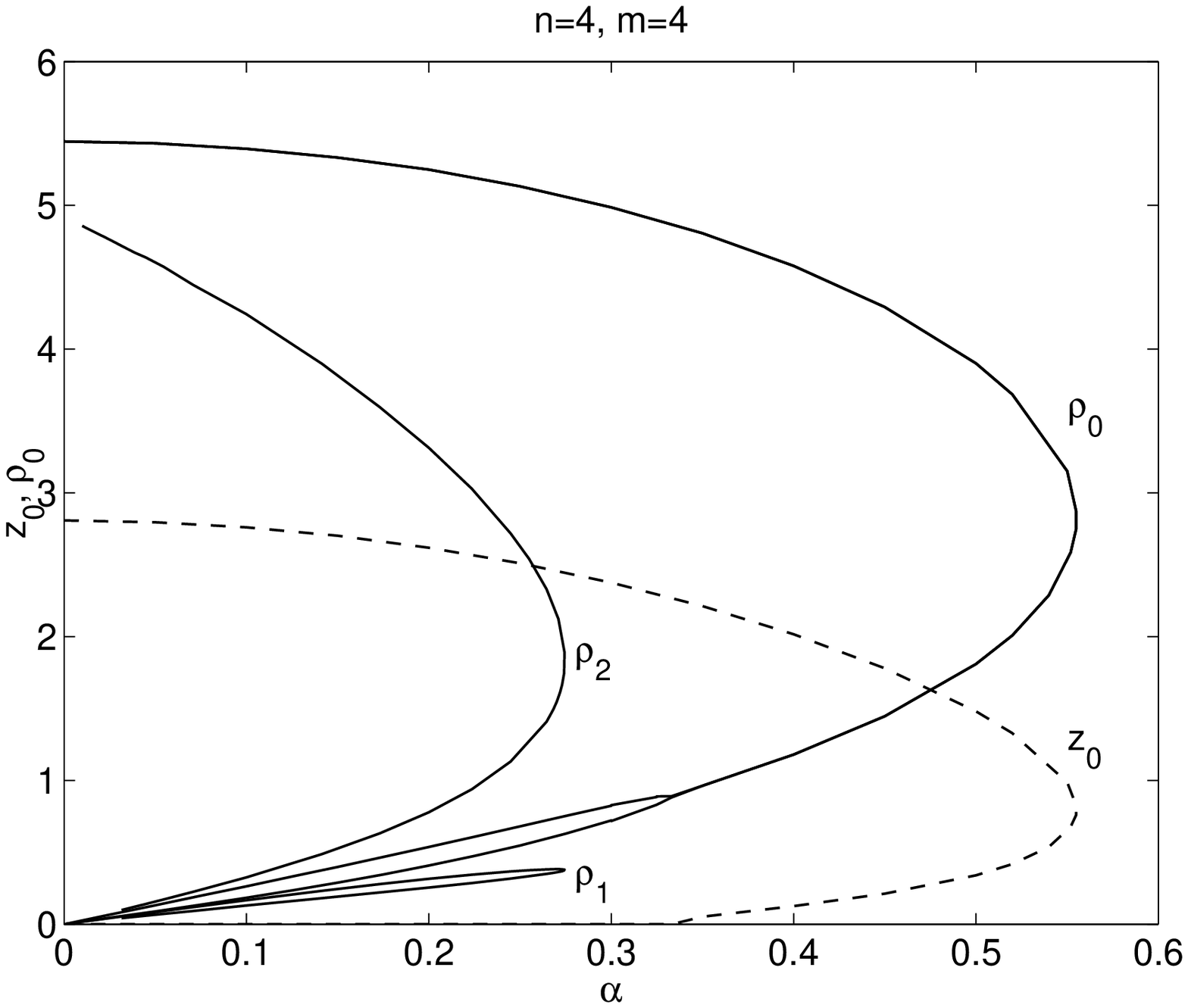} }
(b)\mbox{\epsfysize=6.0cm \epsffile{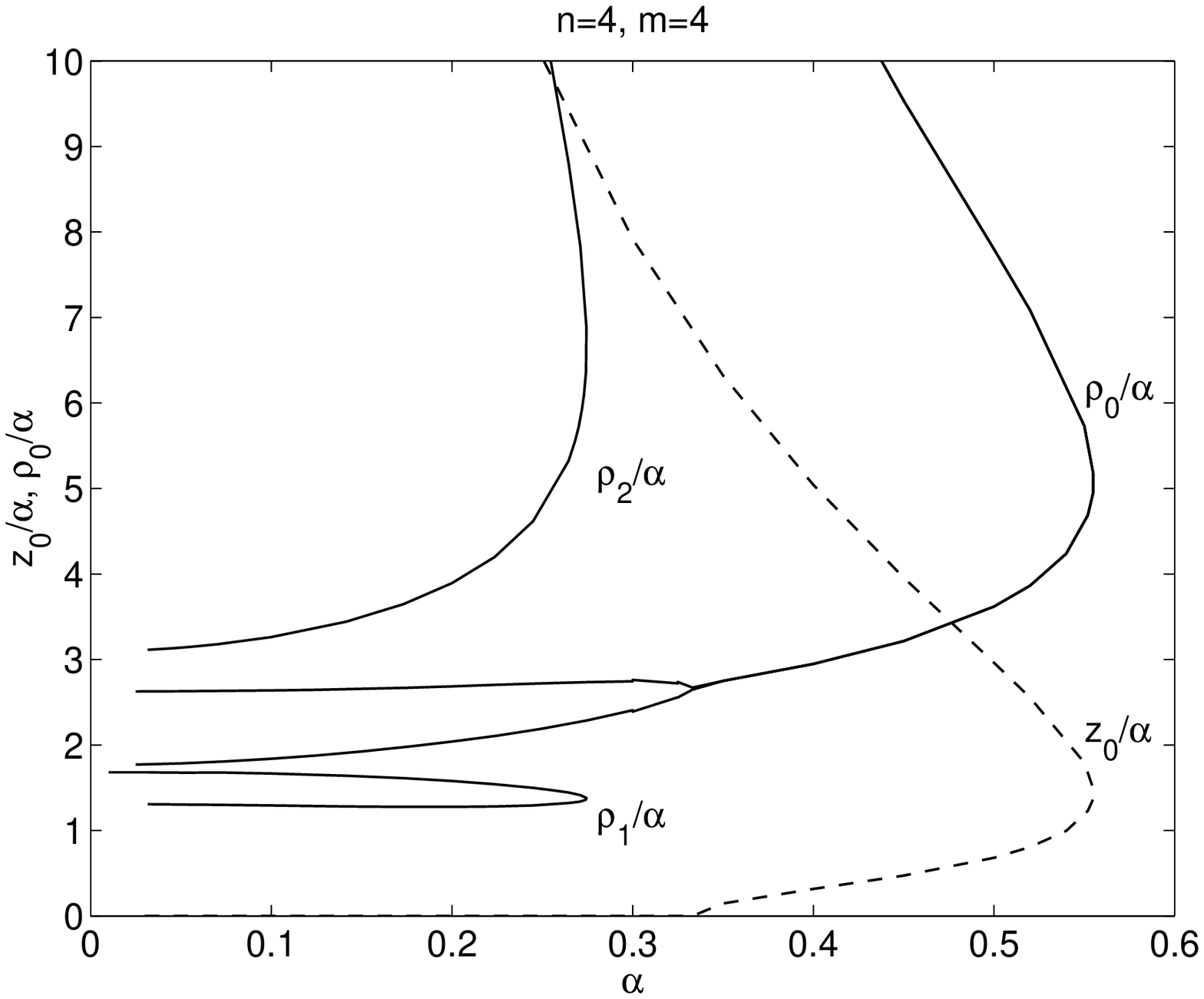} }
}\vspace{0.5cm}
{\bf Fig.~10} \small
The coordinates $\rho$ and $z$ (a) and the scaled coordinates
$\bar \rho$ and $\bar z$ (b) are shown as functions of the
coupling constant $\alpha$
for the $m=4$ and $n=4$ vortex solutions along the four branches.
$\rho_0$, $z_0$ refer to the location of the rings of the first two 
branches, and $\rho_1$, $\rho_2$ to the location of the rings of the 
second two branches.
\vspace{0.5cm}
}

Summarizing, we find the following pattern:
From the flat space $m=4$ and $n=4$ vortex solution
the first lower branch of gravitating vortex solutions
emerges, which merges at the first maximal value $\alpha_{\rm max}$
with the first upper branch.
Along the first upper branch, for small values of $\alpha$ the solutions
approach the scaled $k=2$ and $n=4$ EYM solution with lower mass.
Close to the first upper branch, for small values of $\alpha$ 
the second upper branch arises from 
the scaled $k=2$ and $n=4$ EYM solution with higher mass.
This second upper branch merges at the second smaller 
maximal value $\alpha_{\rm max}$ with the second lower branch
of solutions.
Along the second lower branch, for small values of $\alpha$ the solutions
may be thought of as composed of a scaled
$k=1$ and $n=4$ EYM solution in the inner region
and a flat space $m=2$ and $n=4$ solution in the outer region.

As seen in Fig.~9,
no EYM solutions with $k>2$ and $n=4$ appear.
We therefore conjecture, that for solutions with $m>4$ and $n=4$
an analogous but generalized pattern holds.
There are four branches of solutions.
From the flat space $m>4$ and $n=4$ vortex solution
the first lower branch emerges, 
and merges at a first maximal value $\alpha_{\rm max}$
with the first upper branch.
Along the first upper branch, 
the solutions approach a composite solution for small values of $\alpha$,
consisting of 
the scaled $k=2$ and $n=4$ EYM solution with lower mass in the inner region
and an $m-4$ and $n=4$ YMH solution in the outer region.
The second upper branch arises from a composite solution consisting of
the scaled $k=2$ and $n=4$ EYM solution with higher mass in the inner region
and an $m-4$ and $n=4$ YMH solution in the outer region.
The second upper branch merges at a second 
maximal value $\alpha_{\rm max}$ with the second lower branch
of solutions.
Along the second lower branch, 
for small values of $\alpha$ the solutions may be thought of as
composed of a scaled $k=1$ and $n=4$ EYM solution in the inner region
and a flat space $m-2$ and $n=4$ solution in the outer region.

\boldmath
\subsection{Gravitating vortex rings: $n>4$}
\unboldmath

For vortex solutions with $n=5$ we expect an analogous pattern
as for the $n=4$ vortex solutions.
In particular, there should be four branches of solutions:
a first lower branch emerging from the flat space solution,
a first upper branch approaching a composite solution 
consisting of the lower scaled $k=2$ EYM solution 
and a $m-4$ YMH solution,
a second upper branch arising from a composite solution 
consisting of the higher scaled $k=2$ EYM solution
and a $m-4$ YMH solution,
and a second lower branch approaching a composite solution
consisting of the scaled $k=1$ EYM solution
and a $m-2$ YMH solution.
We illustrate this pattern for the $m=5$ and $n=5$ vortex solutions
in Fig.~11, where we show the gauge field function $K_2$
for solutions on all four branches, for small values of $\alpha$.
The scaled mass along the four branches is shown in Fig.~11c.

\noindent\parbox{\textwidth}{
\centerline{
(a)\mbox{\epsfysize=6.0cm \epsffile{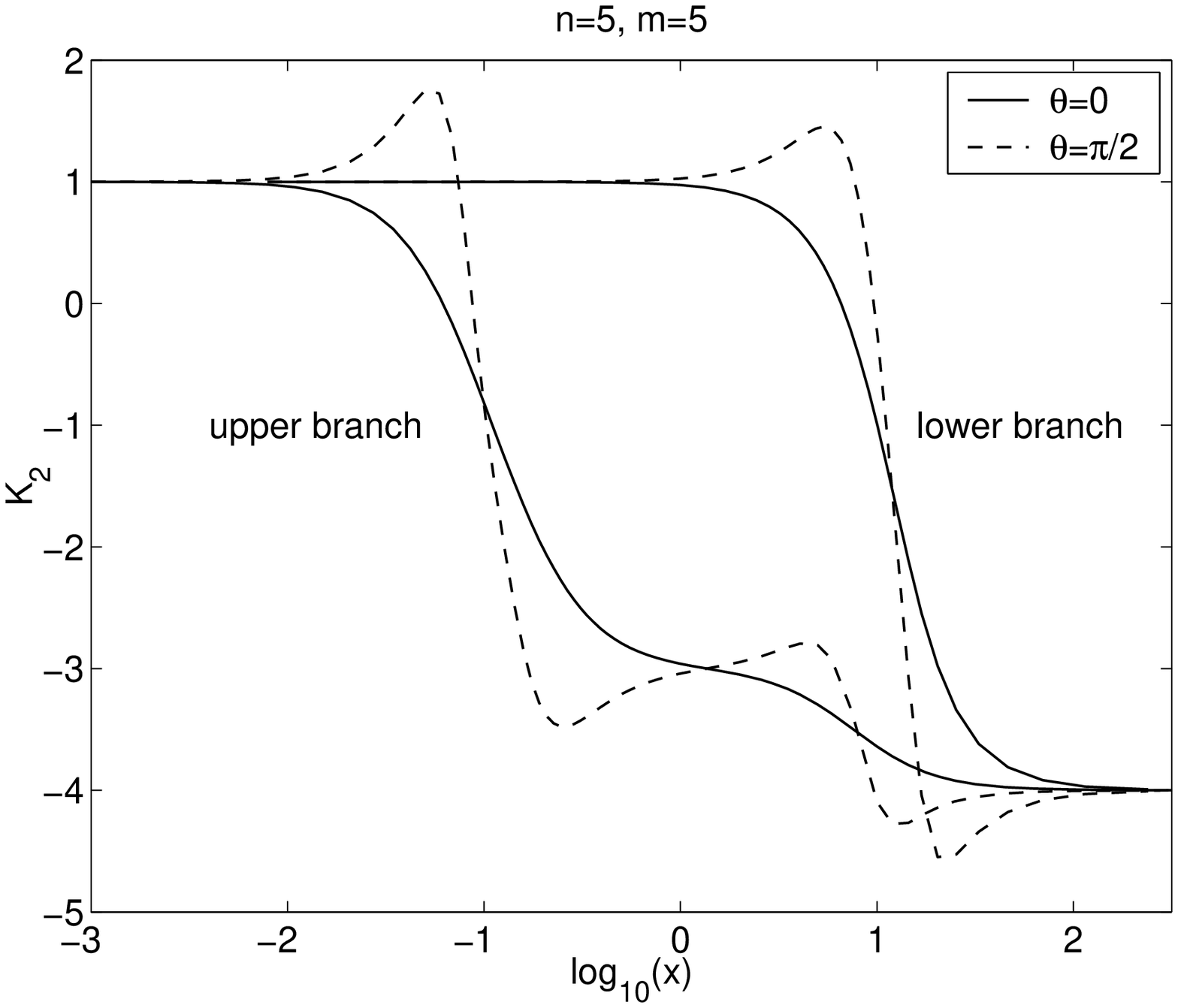} }
(b)\mbox{\epsfysize=6.0cm \epsffile{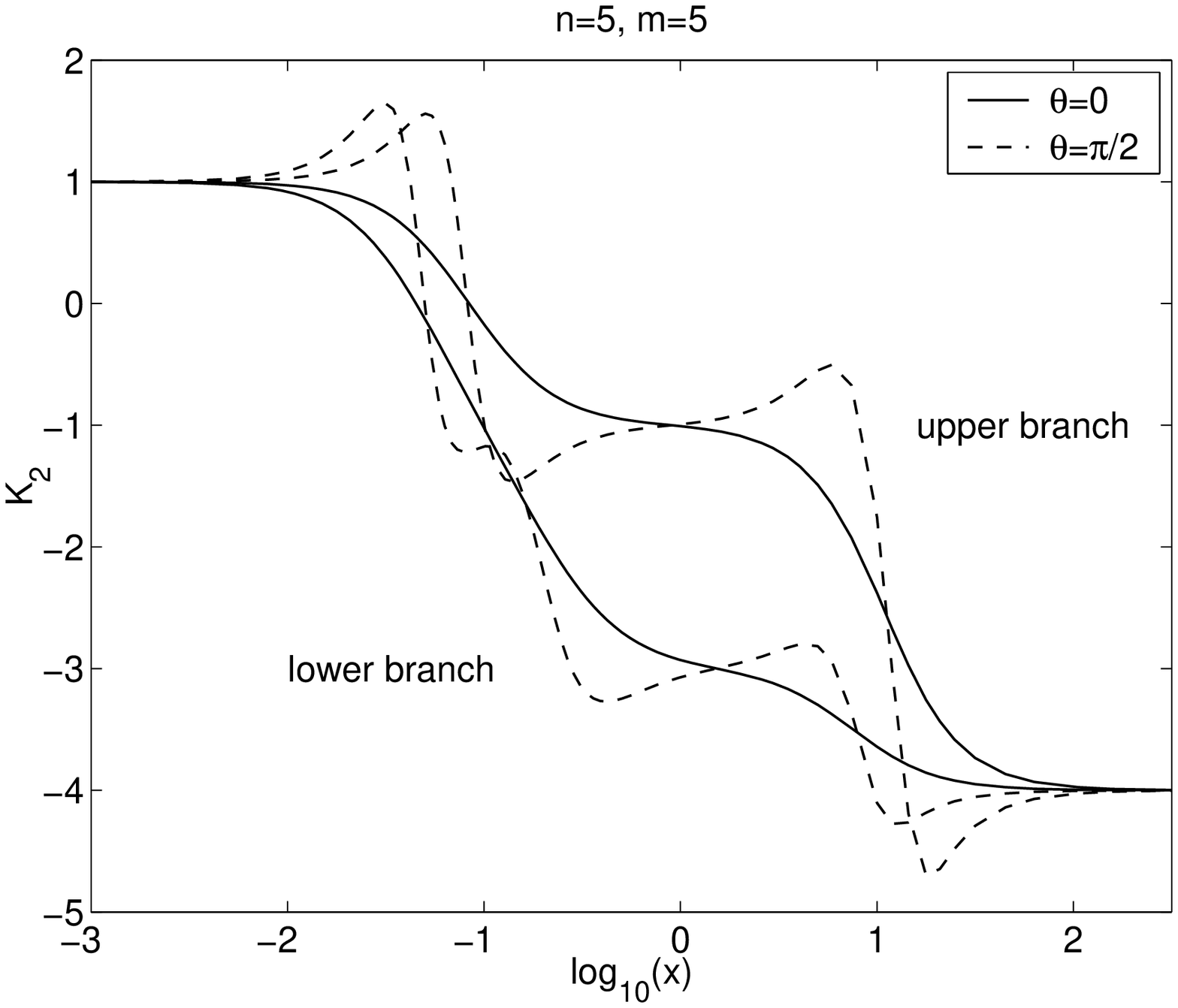} }
}\vspace{0.5cm}
\centerline{
(c)\mbox{\epsfysize=6.0cm \epsffile{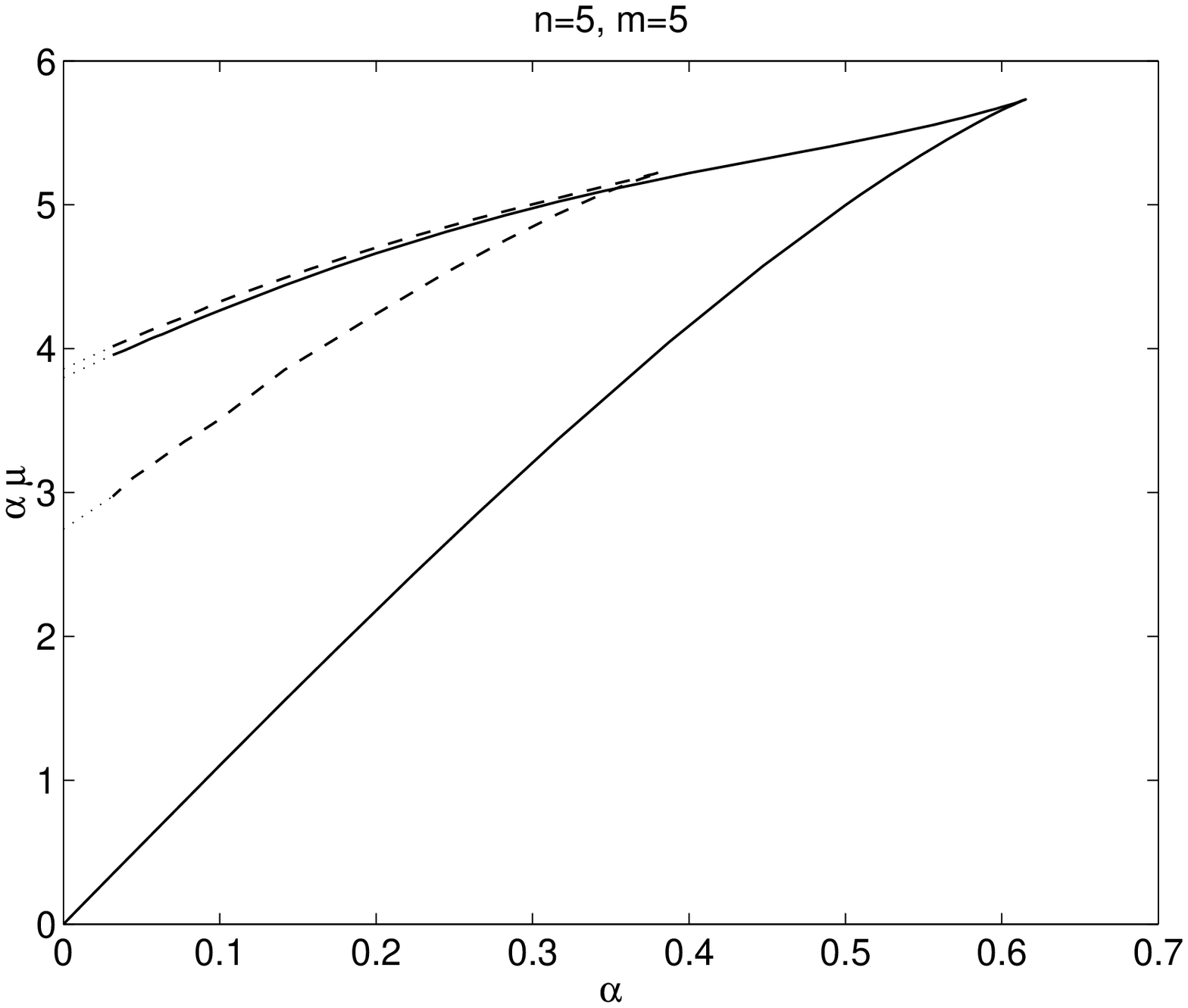} }
}\vspace{0.5cm}
{\bf Fig.~11} \small
The gauge field function $K_2$ 
is shown for the vortex solution with $n=5$ and $m=5$ 
for $\alpha^2=0.001$ on the first lower branch
and the first upper branch (a),
on the second lower branch 
and the second upper branch (b).
The scaled mass $\hat \mu$ 
is shown as function of the coupling constant $\alpha$ 
for all four branches (c).
The dotted lines extend the EYMH curves of the scaled mass to
the masses of the $k=2$ and $k=1$ EYM solutions.
\vspace{0.5cm}
}

Considering higher values of $n$, we observe in Fig.~9,
that for $n=6$ and $m \ge 6$ at least five EYM solutions exist:
one $k=1$, two $k=2$ and two $k=3$ solutions.
This suggests, that for $n=6$ and $m=6$ six branches
of gravitating vortex solutions exist:
a first lower branch emerging from the $n=6$ flat space solution,
a first upper branch approaching the
lower mass scaled $k=3$ and $n=6$ EYM solution,
a second upper branch arising from the
higher mass scaled $k=3$ and $n=6$ EYM solution,
a second lower branch approaching a composite solution
consisting of the lower mass scaled $k=2$ and $n=6$ EYM solution
and a $m-4$ and $n=6$ YMH solution,
a third upper branch arising from a composite solution
consisting of the higher mass scaled $k=2$ and $n=6$ EYM solution
and a $m-4$ and $n=6$ YMH solution,
and a third lower branch approaching a composite solution
consisting of the scaled $k=1$ and $n=6$ EYM solution
and a $m-2$ and $n=6$ YMH solution.
We indeed obtain these six branches numerically.
Surprisingly, there are two more branches for a small range of 
$\alpha$.
We exhibit the scaled mass along these branches
in Fig.~12a. The value of the metric function $l$ at the origin
is exhibited in Fig.~12b.

\noindent\parbox{\textwidth}{
\centerline{
(a) \mbox{\epsfysize=6.0cm \epsffile{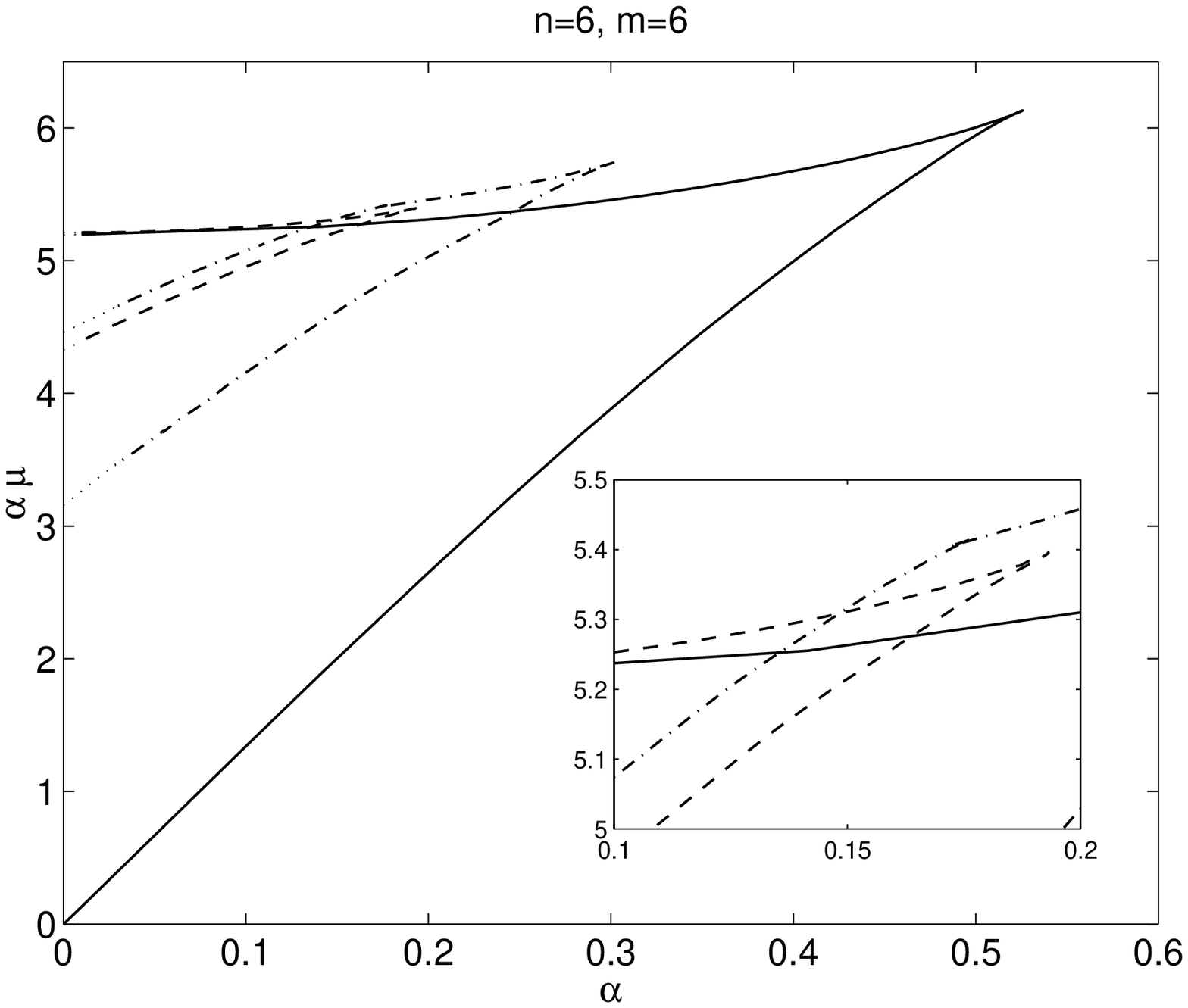} }
(b) \mbox{\epsfysize=6.0cm \epsffile{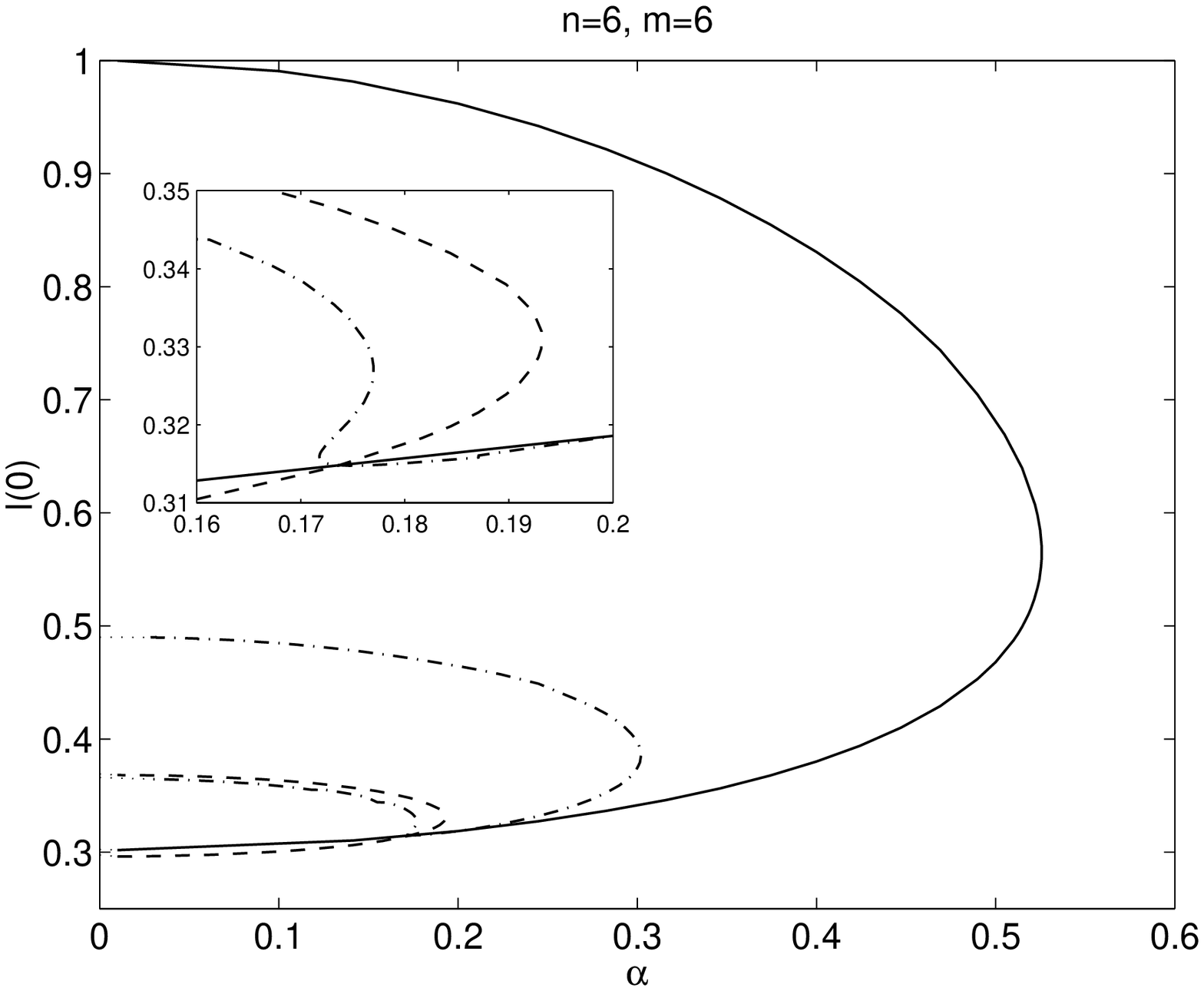} }
}\vspace{0.5cm}
{\bf Fig.~12} \small
The scaled mass $\hat \mu$ (a)
and the value of the metric function $l$ at the origin (b)
are shown as functions of the coupling constant $\alpha$ 
for the six branches of the vortex solutions with $n=6$ and $m=6$.
The dotted lines extend the EYMH curves to
the values of the $k=3$, $k=2$ and $k=1$ EYM solutions.
\vspace{0.5cm}
}

\section{Conclusions}

We have considered the effect of gravity on the
monopole-antimonopole chains and vortex rings
of Yang-Mills-Higgs theory \cite{KKS}.
The resulting solutions are regular, static, axially symmetric 
and asymptotically flat.
They are characterized by two integers $m$ and $n$.
Solutions with $n=1$ and $n=2$ correspond to gravitating chains
of $m$ monopoles and antimonopoles, each of charge $\pm n$, where
the Higgs field vanishes at $m$ isolated points along the symmetry axis.
Larger values of $n$ give rise to gravitating vortex solutions,
where the Higgs field vanishes on one or more rings,
centered around the symmetry axis.

Concerning their dependence on the coupling constant $\alpha$,
we observe the same general pattern for all gravitating
chain solutions.
From the flat space $m$ chain a lower branch of gravitating $m$ chains
emerges, which merges at a maximal value $\alpha_{\rm max}$
with an upper branch.
For small $\alpha$ on the upper branch, the solutions
may be thought of as composed of a scaled (generalized)
BM solution in the inner region and a flat space
$m-2$ solution in the outer region.
Consequently, the mass diverges in the limit $\alpha \rightarrow 0$,
while the scaled mass approaches the mass of the 
lowest (generalized) BM solution of EYM theory.

The same pattern holds for gravitating vortex solutions
with $n=3$, and for $n \ge 4$ when $m \le 3$.
For $n=4$ and $m \ge 4$, however, a new phenomenon occurs.
Instead of two branches of solutions four branches of solutions arise.
The reason for these is the presence of two additional EYM solutions 
\cite{IKKS}. 
In this case we find (and conjecture for higher $m$)
the following pattern for the coupling constant dependence:
A first lower branch emerges from the flat space solution,
a first upper branch approaches a composite solution
consisting of the lower scaled $k=2$ EYM solution
and a $m-4$ YMH solution,
a second upper branch arises from a composite solution
consisting of the higher scaled $k=2$ EYM solution
and a $m-4$ YMH solution,
a second lower branch approaches a composite solution
consisting of the scaled $k=1$ EYM solution
and a $m-2$ YMH solution.

For $n \ge 6$ and $m \ge 6$ not only two but four additional EYM
solutions arise \cite{IKKS}.
We consequently find (and conjecture for higher $m$)
six branches of gravitating vortex solutions, and for a small range
of $\alpha$ even eight branches.
For larger values of $n$ we expect this trend to continue.

We observe that for all the gravitating monopole-antimonpole pair solutions,
chains of monopoles and antimonopoles and vortex rings we considered in this
paper there is no formation of a degenerate horizon, in contrast to 
the gravitating monopole solution \cite{gmono}.

We note that for gravitating Skyrmions \cite{gravSk} the dependence of the 
solutions on the gravitational coupling parameter follows the same pattern as 
for the monopole-antimonpole pair. Again, two branches of solutions 
merge at the maximal value of the coupling parameter, and the upper
branch is connected to the (scaled) lowest mass BM solution.

In this paper we have only considered gravitating chain and vortex solutions
in the limit of vanishing Higgs self-coupling.
For finite values of the self-coupling constant the structure of the
flat space solutions can become more involved, yielding
solutions with vortex rings and several isolated nodes of the Higgs field
\cite{KKS}. Their continuation to curved space remains to be studied.

Furthermore, it would be interesting to consider dyonic gravitating chain 
and vortex solutions \cite{dyon}, as well as black holes with
monopole or dipole hair, consisting of chain-like or vortex-like
structures \cite{dipole}. Finally, recent work on gravitating
monopoles and monopole-antimonopole pairs in the presence of 
a cosmological constant might be generalized to the study of
gravitating chain and vortex solutions \cite{tigran2}.

\begin{acknowledgments}
B.K. gratefully acknowledges support by the DFG under contract 
KU612/9-1.
\end{acknowledgments}

\newpage

\end{document}